\documentclass[11pt,preprint]{aastex}   \usepackage{amssymb,  amsmath}
\usepackage{graphicx} \newcommand{\ud}{\,\mathrm{d}}

\title{Galaxy clustering  and projected density profiles  as traced by
  satellites  in  photometric   surveys:  Methodology  and  luminosity
  dependence}     
\author{
Wenting Wang\altaffilmark{1,2},     
Y.P. Jing\altaffilmark{1},      
Cheng Li\altaffilmark{1},     
Teppei Okumura\altaffilmark{3,1},
Jiaxin Han\altaffilmark{1,2}
}
\altaffiltext{1}{Key Laboratory for Research in Galaxies and Cosmology
of Chinese Academy of Sciences, Max-Panck-Institute Partner Group, 
Shanghai Astronomical Observatory, Nandan Road 80, Shanghai 200030, China}
\altaffiltext{2}{Graduate School  of the Chinese  Academy of Sciences,
 19A, Yuquan Road, Beijing, China} 
\altaffiltext{3}{Institute for the Early Universe, Ewha Womans University, 
Seoul, 120-750, Korea}

\begin{document}

\begin{abstract}
We develop a new method which measures the projected density
distribution $w_p(r_p)n$ of photometric galaxies surrounding a set of
spectroscopically-identified galaxies, and simultaneously the
projected cross-correlation function $w_p(r_p)$ between the two
populations.  In this method we are able to divide the photometric
galaxies into subsamples in luminosity intervals even when redshift
information is unavailable, enabling us to measure $w_p(r_p)n$ and
$w_p(r_p)$ as a function of not only the luminosity of the
spectroscopic galaxy, but also that of the photometric
galaxy. Extensive tests show that our method can measure $w_p(r_p)$ in
a statistically unbiased way. The accuracy of the measurement depends
on the validity of the assumption inherent to the method that the
foreground/background galaxies are randomly distributed and are thus
uncorrelated with those galaxies of interest. Therefore, our method can be
applied to the cases where foreground/background galaxies are
distributed in large volumes, which is usually valid in real
observations.

 We have applied our method to data from the Sloan Digital Sky Survey
 (SDSS) including a sample of $10^5$ luminous red galaxies (LRGs) at
 $z\sim0.4$ and a sample of about half a million galaxies at
 $z\sim0.1$, both of which are cross-correlated with a deep
 photometric sample drawn from the SDSS.  On large scales, the
 relative bias factor of galaxies measured from $w_p(r_p)$ at
 $z\sim0.4$ depends on luminosity in a manner similar to what is found
 for those at $z\sim0.1$, which are usually probed by autocorrelations
 of spectroscopic samples in previous studies.  On scales smaller than
 a few Mpc and at both $z\sim0.4$ and $z\sim0.1$, the photometric
 galaxies of different luminosities exhibit similar density profiles
 around spectroscopic galaxies at fixed luminosity and redshift. This
 provides clear observational support for the assumption
 commonly-adopted in halo occupation distribution (HOD) models that
 satellite galaxies of different luminosities are distributed in a
 similar way, following the dark matter distribution within their host
 halos.

\end{abstract}

\section{INTRODUCTION}

In cold  dark matter dominated cosmological models,  dark matter halos
form in density peaks in  the universe under the influence of gravity,
and thus  are clustered  in a different  way from the  underlying dark
matter.  In other words, they are {\em biased} in spatial distribution
relative    to    dark   matter    \citep[e.g.][]{1996MNRAS.282..347M,
  1998ApJ...503L...9J, 2004MNRAS.355..129S}.  Galaxies are believed to
form inside these  halos \citep[]{1978MNRAS.183..341W}, and thus their
spatial  distribution  is also  biased  with  respect  to dark  matter
\citep[e.g.][]{1984ApJ...284L...9K,                1985ApJ...292..371D,
  1986ApJ...304...15B}.  On  large scales ($\ga  10$Mpc), such biasing
is nearly linear  and the clustering of dark  matter is well described
by  linear  perturbation  theory.   On smaller  scales,  in  contrast,
galaxies  do  not  trace  dark matter  simply.   Complicated  physical
processes  involved  in galaxy  formation  and  evolution  have to  be
considered  if   one  desires  to  fully   understand  galaxy  clustering
\citep[e.g.][]{1991ApJ...379...52W,                1999MNRAS.303..188K,
  2000MNRAS.319..209C}.  This  leads galaxy clustering  and biasing to
depend on a  variety of factors including spatial  scale, redshift and
galaxy properties.  Therefore measuring  the clustering of galaxies as
a function of  their physical properties over large  ranges in spatial
scale  and redshift  is helpful  for understanding  how  galaxies have
formed and evolved.

Recent  large redshift  surveys, in  particular the  Two  Degree Field
Galaxy  Redshift  Survey  \citep[2dFGRS]{2001MNRAS.328.1039C} and  the
Sloan  Digital   Sky  Survey  \citep[SDSS]{2000AJ....120.1579Y},  have
enabled detailed studies on  galaxy clustering in the nearby universe.
These studies  have well established  that the clustering  of galaxies
depends on a variety of  properties, such as luminosity, stellar mass,
color,  spectral  type,  and morphology  \citep[]{2001MNRAS.328...64N,
  2002MNRAS.332..827N,    2003MNRAS.344..847M,    2002ApJ...571..172Z,
  2005ApJ...630....1Z,    2003MNRAS.346..601G,    2006MNRAS.368...21L,
  2010arXiv1005.2413T}.  More luminous (massive) galaxies are found to
cluster more strongly than  less luminous (massive) galaxies, with the
luminosity  (mass)  dependence  being  more  remarkable  for  galaxies
brighter  than  $L_\ast$  (the  characteristic  luminosity  of  galaxy
luminosity function described by  a Schechter function, \cite{1976ApJ...203..297S}).
  Moreover, galaxies  with redder
colors, older  stellar populations and  more bulge-dominated structure
show  higher  clustering  amplitudes   and  steeper  slopes  in  their
two-point correlation functions.

There have  also been  recent studies on  galaxy clustering  at higher
redshifts.    At  $z\sim   1$,  the   DEEP2  Galaxy   Redshift  Survey
\citep[]{2003SPIE.4834..161D}   and    the   VIMOS-VLT   Deep   Survey
\citep[VVDS]{2005A&A...439..845L}  have shown  that  galaxy clustering
depends  on  luminosity,  stellar   mass,  color,  spectral  type  and
morphology,  largely consistent  with  what are  found  for the  local
universe        \citep{2004ApJ...609..525C,       2006ApJ...644..671C,
  2006A&A...452..387M,    2008ApJ...672..153C,    2008A&A...478..299M,
  2009arXiv0911.2252D}.     In    contrast,    the   zCOSMOS    survey
\citep[]{2007ApJS..172...70L} shows no  clear luminosity dependence of
galaxy  clustering  over  redshift  range  $0.2\leqslant  z  \leqslant
1$\citep{2009A&A...505..463M}.    More  surprisingly,   the  projected
two-point  auto-correlation  function   $w_p(r_p)$  derived  from  the
zCOSMOS is significantly higher and flatter than from the VVDS \citep{
  2008A&A...478..299M, 2009A&A...505..463M}.

The observational  measurements of galaxy clustering  at both $z\sim0$
and $z\sim1$ as described above have been widely used to test theories
of                           galaxy                          formation
\citep[e.g.][]{1997MNRAS.286..795K,2000MNRAS.316..107B,
  2007MNRAS.376..984L,2010MNRAS.404.1111G}  , as  well as  to quantify
the  evolution  of  galaxy  clustering  from  high  to  low  redshifts
\citep[e.g.][]{2007ApJ...667..760Z,2008A&A...478..299M,
  2010MNRAS.402.1796W}.   Galaxy  clustering  has  also been  used  to
constrain  halo  occupation  distribution(HOD) models,  which  provide
statistical description on how galaxies are linked to their host halos
and   hence   useful   clues   for   understanding   galaxy   formation
\citep[e.g.,][]{1998ApJ...494....1J,               1998ApJ...503...37J,
  2000MNRAS.318.1144P,2000ApJ...543..503M,         2000MNRAS.318..203S,
  2001ApJ...546...20S,                             2002ApJ...575..587B,
  2002PhR...372....1C,2003MNRAS.339.1057Y,         2005ApJ...633..791Z,
  2005ApJ...631...41T}.

At  intermediate  redshifts ($0.2\la  z\la1$),  progress on  measuring
galaxy clustering has been relatively hampered by the lack of suitable
data   sets.    A   few  studies   \citep[e.g.][]{2001ApJ...560...72S,
  2001ApJ...563..736C,  2002MNRAS.332..617F, 2006A&A...457..145P} have
measured galaxy clustering  as a function of color  which are in broad
agreement  with results found  for the  local universe.   However, the
dependence  of clustering  on luminosity,  which is  well seen  in the
local universe,  has not been fully established  at these intermediate
redshifts, very  likely due to  the limited size of  the spectroscopic
samples. These samples usually cover  small area on the sky, suffering
from  both  sampling  noise   and  large-scale  structure  noise  (the
so-called {\em cosmic variance} effect).

In  this paper, rather  than measuring  the {\em  auto-correlation} of
these galaxies  as in most previous  studies, we develop  a new method
for   estimating  the   projected  two-point   {\em  cross-correlation
  function}  $w_p(r_p)$  between  a  given  set  of  spectroscopically
identified galaxies  and a large  sample of photometric  galaxies.  In
brief,  we  first  estimate  the  angular  cross-correlation  function
between  the  spectroscopic  and  the  photometric  samples.  We  then
determine   the  projected,   average   number  density   distribution
$w_p(r_p)n$ of the  photometric galaxies surrounding the spectroscopic
objects, as well as the projected two-point cross-correlation function
$w_p(r_p)$.  The  photometric sample is  usually the parent  sample of
the  spectroscopic  galaxies,  but   goes  to  much  fainter  limiting
magnitudes.  The spectroscopic sample could be clusters (or groups) of
galaxies, central galaxies  of dark matter halos such  as the luminous
red galaxies (LRGs) in the  SDSS, quasars, or any spectroscopic galaxy
populations of  interest.  Our method  can yield a measurement  of the
projected  density  distribution  of  galaxies with  certain  physical
properties  (such  as luminosity,  color,  etc.) around  spectroscopic
objects of  certain properties. In  this paper we focus  on presenting
our  methodology and limit  the application  to galaxies  of different
luminosities.  We plan to  examine the  dependence of  $w_p(r_p)n$ and
$w_p(r_p)$ on  other properties (color, morphology,  etc.)  in future
work.

Previous studies of satellite galaxy distribution around relatively
bright galaxies are mostly limited to low redshifts \citep[$z<0.1$,
  e.g.]  []{1980ApJ...238L..13L, 1987MNRAS.229..621P,
  1991ApJ...379L...1V, 1994MNRAS.269..696L, 2005MNRAS.356.1045S,
  2006ApJ...647...86C}.  \cite{2006ApJ...644...54M} and
\cite{2005ApJ...621...22Z} have recently investigated
cross-correlations between spectroscopic and imaging galaxy samples at
intermediate redshift ($0.2\lesssim z \lesssim 0.4$), but with
different methods and focuses.  Here we apply our method to a deep,
photometric galaxy catalogue and a spectroscopic LRG sample at
$z\sim0.4$, both of which are drawn from the final data release of the
SDSS \citep[]{2009ApJS..182..543A} . The LRGs are expected to be the
central galaxy of their host dark matter halos.  Therefore, by
measuring $w_p(r_p)n$ on scales smaller than a few Mpc, we yield an
estimate of the density distribution of satellites galaxies within
their host halo, as well as its dependence on luminosities of both
central and satellite galaxies.  On larger scales, our analysis leads
to a measurement of linear relative bias factor for photometric
galaxies of different luminosities.

We describe  our galaxy samples  in \S~\ref{sec:data} and  present our
methodology  in \S~\ref{sec:method}.   Applications to  SDSS  data are
presented in \S~\ref{sec:apply to  sdss}.  We summarize and discuss in
the last  section.  Throughout this  paper we assume a  cosmology with
$\Omega_m=0.3$,$\Omega_\Lambda=0.7$   and    $H_0=100h   $km$   s^{-1}
$Mpc$^{-1}$ ($h=1$).

\section{Data}
\label{sec:data}

\subsection{The LRG sample at intermediate redshift}

\begin{figure}
\label{fig:dndz_lrg}\epsscale{0.9}\plotone{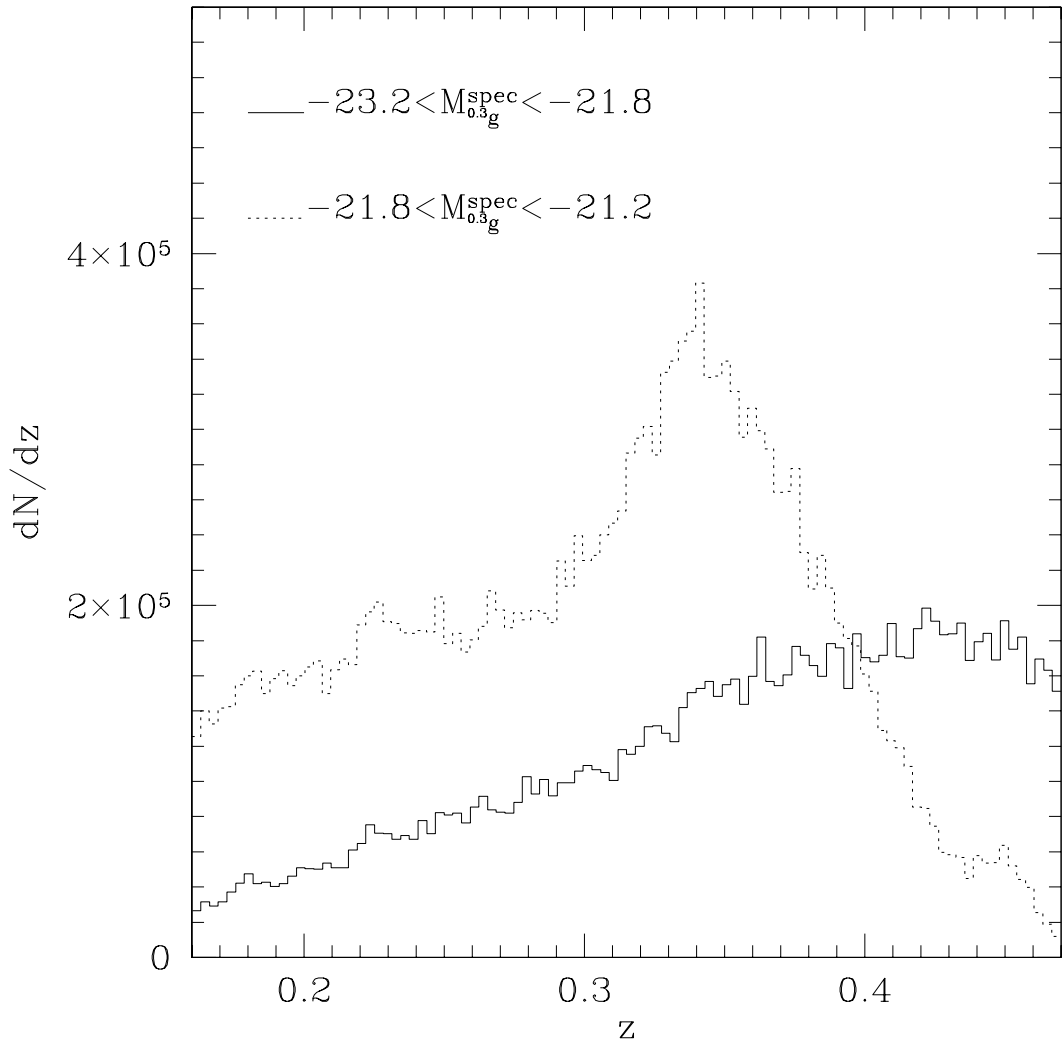}
\caption{Redshift histograms for luminous red galaxies (LRGs) in our 
sample falling in the two luminosity intervals, as indicated, which we 
use to select our subsamples in Table 1. The g-band absolute magnitude
  $M_{^{0.3}g}$ is $K-$ and $E-$ corrected to its value at $z=0.3$. 
}
\end{figure}

The LRG sample is constructed from the SDSS data release 7
\citep[DR7][]{2009ApJS..182..543A}, consisting of 101,658 objects with
spectroscopically measured redshift in the range $0.16<z<0.47$,
absolute magnitude limited to $-23.2<M_{^{0.3}g}<-21.2$ and redshift
confidence parameter greater than 0.95.  Here $M_{^{0.3}g}$ is the
$g$-band absolute magnitude $K$- and $E$-corrected to redshift $z=0.3$
(see \citealt{2001AJ....122.2267E} for references).  We further select
those LRGs that are expected to be the central galaxy of their host
dark matter halos, using a method similar to that adopted in
\cite{2009ApJ...698..143R} and \cite{2009ApJ...694..214O}.  We use
linking lengths of 0.8 $h^{-1}$Mpc and 20 $h^{-1}$Mpc for separations
perpendicular and parallel to the line of sight when linking galaxies
into groups.  This leads to a total of 93802 central galaxies (about
$92.3\%$ of the initial LRG catalogue), covering a sky area which is
almost the same as that of \cite{2010ApJ...710.1444K}.  From this
catalogue we select five samples in two luminosity intervals
($-23.2<M_{^{0.3}g}<-21.8$ and $-21.8<M_{^{0.3}g}<-21.2$) and in three
redshift intervals ($0.16<z<0.26$, $0.26<z<0.36$ and
$0.36<z<0.46$). Details of our samples are listed in
Table~\ref{tab:lrg}.  These samples so selected are volume limited,
except {\tt Sample L4} which is approximately, but not perfectly
volume limited as can be seen from fig.  1 of
\cite{2005ApJ...621...22Z}.  Figure~\ref{fig:dndz_lrg} shows the
redshift distributions of our LRGs in the two luminosity intervals.

\subsection{The low-redshift galaxy sample}

\begin{figure}
\label{fig:dndz_lowz}\epsscale{0.9}\plotone{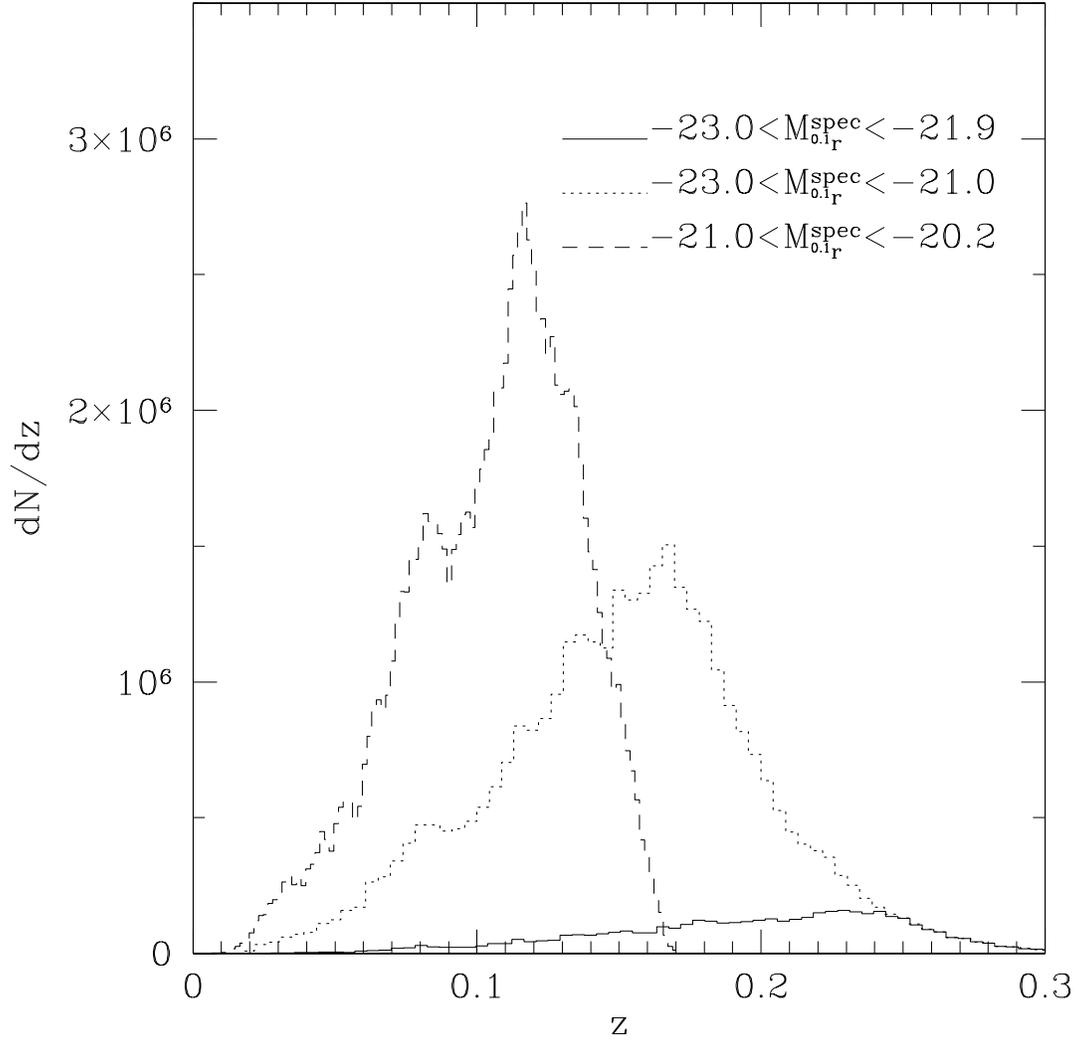}
\caption{Redshift histograms for spectroscopic galaxies in our low-redshift
sample in three different luminosity ranges as indicated, which we use
to select our subsamples in Table 2.  The $r$-band absolute magnitude 
$M_{^{0.1}g}$ is $K-$ and $E-$ corrected to its value at $z=0.1$.}
\end{figure}

Our spectroscopic  galaxy sample in the local  Universe is constructed
from   the   New   York   University  Value   Added   Galaxy   Catalog
(NYU-VAGC)\footnote{http://sdss.physics.nyu.edu/vagc/}, which is built
by \cite{2005AJ....129.2562B} based on the SDSS DR7. From the NYU-VAGC
we  select   a  magnitude-limited  sample  of   533,731  objects  with
$0.001<z<0.5$   and  $r$-band   Petrosian  magnitude   in   the  range
$10.0<r<17.6$.  The sample has a median redshift of $z=0.09$, with the
majority of the  galaxies at $z<0.25$.  The galaxies  are divided into
five non-overlapping redshift bins,  ranging from $z=0.03$ to $z=0.23$
with  an equal  interval of  $\Delta{z}=0.04$.  The  galaxies  in each
redshift  bin are  further  restricted to  various luminosity  ranges,
giving rise  to a  set of  eight volume limited  samples as  listed in
Table~\ref{tab:lowz}.   The $r$-band absolute  magnitude $M_{^{0.1}r}$
is  $K$- and  $E$-corrected to  its  value at  $z=0.1$ following
\cite{2003ApJ...592..819B} (hereafter B03).
Figure~\ref{fig:dndz_lowz}  shows  the  redshift distribution  of  the
galaxies falling  into the three  luminosity ranges which are  used to
select  the  samples.  These  samples  by  construction  are at  lower
redshifts when compared to the LRG samples, allowing us to make use of
photometric galaxies  for our analysis over wide  ranges in luminosity
and redshift.  We don't attempt to select central galaxies as done for
LRGs above,  as it is  not straightforward to  do so. Thus  one should
keep in mind  that by using the low-redshift  samples selected here we
will measure  density profiles and projected  correlations for general
populations of galaxies, not only for central galaxies.

\subsection{The photometric galaxy sample and random samples}

We construct our photometric galaxy sample from the {\tt datasweep}
catalogue which is included as a part of the NYU-VAGC.  This is a
compressed version of the full photometric catalogue of the SDSS DR7
that was used by \cite{2005AJ....129.2562B} to build the NYU-VAGC.  It
contains only decent detections and includes a subset of all
photometric quantities, which is enough for our analysis.  Starting
from the datasweep catalogue, we select all galaxies with $r$-band
apparent Petrosian magnitudes in the range $10 < r < 21$ after a
correction for Galactic extinction and with point spread function and
model fluxes satisfying $f_{model} > 0.875 \times f_{PSF}$ in all five
bands.  In order to select unique objects in a run that are not at the
edge of the field, we require the {\tt RUN PRIMARY} flag to be set and
the {\tt RUN EDGE} flag not to be set. Finally we also require the
galaxies to be located within {\em target tiles} of the Legacy Survey
\citep[]{2003AJ....125.2276B}.  This procedure results in a sample of
$\sim21.1$ million galaxies.

As shown by  \cite{2007ApJ...665...67R}, the datasweep catalogue needs
to  be properly  masked,  otherwise the  angular correlation  function
obtained would be falsely flat  on large scales.  We describe the SDSS
imaging   geometry   in   terms   of   disjoint   spherical   polygons
\citep[]{2002MNRAS.330..506H,                      2002MNRAS.335..887T,
  2005AJ....129.2562B}, which  accompany the NYU-VAGC  release and are
available from the NYU-VAGC website. We exclude all polygons (and thus
the galaxies located within them) which contain any object with seeing
greater than $1.5^{\prime\prime}$ or Galactic extinction $A_r>0.2$. We
also  exclude polygons  that intersect  the  mask for  galaxy M101  as
described in  \cite{2007ApJ...665...67R}.  As a result,  a fraction of
about $14.32\%$ of the total  survey area has been discarded, slightly
larger  than in \cite{2007ApJ...665...67R}  where the  authors exclude
image  pixels rather than  polygons with  less critical  criteria than
adopted here.  We  also restrict ourselves to galaxies  located in the
main  contiguous area  of the  survey  in the  northern Galactic  cap,
excluding the  three survey strips in  the southern cap  (about 10 per
cent of the  full survey area).  These restrictions  result in a final
sample of $\sim19.7$ million galaxies.

We  have  constructed a  random  sample  which  has exactly  the  same
geometry and limiting magnitudes as the real photometric sample.  This
is done by generating sky positions  (RA and Dec) at random within the
polygons covering the real galaxies. In this work both the photometric
sample  and  the  random  sample  are cross-correlated  with  a  given
spectroscopic    sample   to    estimate    the   two-point    angular
cross-correlation function  $w(\theta)$ between the  spectroscopic and
the photometric samples.

\section{Methodology}
\label{sec:method}

In our method we estimate in the first place the angular
cross-correlation function $w(\theta)$ between a given set of
spectroscopic galaxies selected by luminosity and redshift (samples
listed in Tables~\ref{tab:lrg} and ~\ref{tab:lowz}) and a set of
photometric galaxies that, if they were at the redshift of the
spectroscopic sample, would be expected to fall in a given luminosity
range.  Next, we convert $w(\theta)$ to determine the projected
density distribution $w_p(r_p)n$ of the photometric galaxies around
the spectroscopic galaxies, from which we further estimate the
projected cross-correlation function $w_p(r_p)$ between the two
populations.  In this section we describe how we select photometric
galaxies in a given luminosity range, followed by description of our
measures of $w(\theta)$ as well as the way of determining $w_p(r_p)n$
and $w_p(r_p)$.

\subsection{Selecting photometric galaxies according to luminosity and redshift}
\label{sec:selecting_sat}

Considering a sample of spectroscopic galaxies with absolute magnitude
and redshift in the ranges $M_{s,1}<M_{s}<M_{s,2}$ and
$z_1<z_{s}<z_2$, we want to measure the cross-correlation of this
sample with a set of photometric galaxies with absolute magnitude
$M_{p,1}<M_{p}<M_{p,2}$.  Due to the lack of redshift information for
the photometric sample, it is not straightforward to determine which
galaxies should be selected in order to have a subset falling in the
expected luminosity range. One can overcome this difficulty by the
fact that the cross-correlation signal is dominated by those
photometric galaxies that are at the same redshifts as the
spectroscopic objects, while both foreground (below $z_1$) and
background (above $z_2$) galaxies contribute little.  This is
reasonably true when the redshift interval $z_2-z_1$ and the projected
physical separation $r_p$ in consideration are substantially small,
for the clustering power decreases rapidly (approximately a power law)
with increasing separation. On large scales, projection effect due to
contamination of foreground and background galaxies becomes relatively
large (we will discuss more about this point in
\S~\ref{sec:directconv}).  With this assumption in mind we restrict
ourselves to photometric galaxies with apparent magnitude
$m_{p,1}<m_{p}<m_{p,2}$, the magnitude range for photometric galaxies
to have absolute magnitude in the range $M_{p,1}<M_{p}<M_{p,2}$ at
redshift $z_1<z_{p}<z_2$, when estimating angular cross-correlation
functions.

When calculating the apparent magnitude for a given absolute magnitude
and redshift, we have  adopted the empirical formula of $K$-correction
presented by \cite{2010arXiv1006.2823W}, which  works at $r$-band as a
function of  observed $g-r$ color and  redshift.  Since $K$-correction
value  changes slowly  with redshift,  we adopt  $(z_1+z_2)/2$  as the
input  redshift value when  applying the  formula for  simplicity.  In
this paper  we adopt the $^{0.1}r$-band luminosity  function from B03,
and so we convert the apparent magnitude in $r$ to that in ${^{0.1}r}$
using    the    analytical     conversion    formula    provided    by
\cite{2007AJ....133..734B}.

\subsection{Measuring angular correlation functions}

We  use  two  estimators  to  measure the  angular  cross  correlation
function between a spectroscopic sample and a photometric sample.  For
small  separations   ($\theta\la  1000^{\prime\prime}$)  we   use  the
standard estimator \citep[]{1983ApJ...267..465D}
\begin{equation}\label{eqn:standard_estimator}
 w(\theta)= \frac{QD(\theta)}{QR(\theta)}-1,
\end{equation}
where  $\theta$  is  the  angular  separation,  and  $QD(\theta)$  and
$QR(\theta)$  are  the cross  pair  counts  between the  spectroscopic
sample and the photometric  sample, and between the same spectroscopic
sample and the random sample.   Note that $QR$ is normalized according
to the ratio  of the size of the photometric  and random samples.  For
separations  larger than  $\theta\sim 1000\arcsec$,  we instead  use a
Hamilton-like estimator \citep[]{1993ApJ...417...19H}
\begin{equation}\label{eqn:hamilton_estimator}
 w(\theta)= \frac{QD(\theta)RR(\theta)}{QR(\theta)DR(\theta)}-1,
\end{equation}
where $RR$ is the pair count  of the random sample, and $DR$ the cross
pair count between the photometric and random samples.  This estimator
is  expected to work  better on  large scales  than the  standard one,
since it is less sensitive to uncertainties in the mean number density
of   photometric   galaxies   \citep{1993ApJ...417...19H}.   The   two
estimators differ in  $w(\theta)$ by $10\%$ to $20\%$  at $\theta \sim
1000\arcsec$ in our  cases. On smaller scales the  two estimators give
almost  identical results, with  difference at  a few  percent level  and well
within error bars.  In order to reduce computation  time, we apply the
standard  estimator  to a  random  sample  of  30 million  points  for
separations $\theta  \lesssim 1000\arcsec$, while a  smaller sample of
$0.9$ million  random points and the Hamilton-like  estimator are used
for larger separations.

\subsection{Converting $w(\theta)$ to $w_p(r_p)n$ and $w(r_p)$}

Given  a  measurement of  $w(\theta)$  we  estimate the  corresponding
projected cross-correlation function  $w_p(r_p)$ and projected density
profile $w_p(r_p)n$ in the following  two ways. In this subsection the
photometric  and  spectroscopic galaxy  samples  being considered  are
named {\tt Sample 1} and {\tt Sample 2}, respectively.

\subsubsection{Direct conversion from $w(\theta)$ to $w_p(r_p)$}
\label{sec:directconv}

The relation between angular correlation function $w(\theta)$ and real
space correlation function $\xi(r)$ is given by (see Peebles 1980)
\begin{equation}\label{eqn:wtheta_xir}
w(\theta)=\frac {\int_{0}^{\infty} {x_1^2 x_2^2  \ud x_1 \ud x_2 a_1^3
    a_2^3 n_1 n_2 \xi(r_{1,2})}}{{\cal{N}}_1{\cal{N}}_2},
\end{equation}
where  $a$ and  $x$  stands  for scale  factor  and comoving  distance
respectively; $r_{1,2}$ is the  real space separation between Sample 1
and Sample 2 galaxies; ${\cal{N}}_1$ and ${\cal{N}}_2$ are the surface
number  densities of  the two samples;  $n_1$ and  $n_2$  are their comoving
spatial  number densities.  Taking  $n_2$ as  a
sum of Dirac delta functions, we have
\begin{equation}
n_2=\sum_k {\delta(\vec{r_2}-\vec{r_k})},
\end{equation}
where  $\vec{r_k}$  stands for  galaxy positions  in Sample  2.  Thus
Eqn.~(\ref{eqn:wtheta_xir}) becomes
\begin{eqnarray}
w(\theta)  &  =  &  \frac {\Omega_1  \Omega_2  \sum_k\int_{0}^{\infty}
  {x_1^2    x_2^2    \ud    x_1    \ud    x_2    a_1^3    a_2^3    n_1
    \delta(\vec{r_2}-\vec{r_k}) \xi(r_{1,2})}}{N_1 N_2}\\  & = & \frac
{\Omega_1   \sum_k\int_{0}^{\infty}   {x_1^2   \ud   x_1   a_1^3   n_1
    \xi(r_{1,k})}}{N_1  N_2}\\ & =  & \frac  { \sum_k\int_{0}^{\infty}
  {\ud        N_1/\ud       z_1        \xi(r_{1,k})\ud       z_1}}{N_1
  N_2},\label{eqn:simplified_wtheta_xir}
\end{eqnarray}
where
\begin{equation}
r_{1,k}=\sqrt{r_1^2+r_k^2-2r_1r_kcos(\theta)},
\end{equation}
and $r_1$ and $r_k$ are  the comoving distances for galaxies in Sample
1 and the $k$th galaxy in Sample 2. Here $N_1$ and $N_2$ are the total
number of objects in the two  samples.  Let $N_k$ denote the number of
galaxies  with  approximately the  same  distance  $r_k$ (or  redshift
$z_k$) in Sample 2, then Eqn.~(\ref{eqn:simplified_wtheta_xir}) can be
written as
\begin{eqnarray}\label{eqn:simplified_wtheta_xir2}
  w(\theta)=  \frac  {\sum_k N_k  \int_{0}^{\infty}  {\ud N_1/\ud  z_1
      \xi(r_{1,k})\ud z_1}}{N_1 N_2}.
\end{eqnarray}
If  the redshift  bin of  Sample 2  is thin  enough, all  the galaxies
within it can be regarded as  at the same redshift. This gives rise to
a much simplified relation between $w(\theta)$ and $\xi(r)$:
\begin{equation}\label{eqn:final_wtheta_xir}
  w(\theta)=  \frac {\int_{0}^{\infty}  {\ud N_1/\ud  z_1 \xi(r_{1,2})
      \ud z_1}}{N_1}.
\end{equation}

On the  other hand,  the relation between  $w_p(r_p)$ and  $\xi(r)$ is
known to be
\begin{equation}\label{eqn:wrp_xir}
  w_p(r_p)=2        \int_{r_p}^{\infty}       {\xi(r)\frac{r       \ud
      r}{\sqrt{r^2-r_p^2}}}  = \int_{z_l}^{z_u} {\xi(r_{1,2})\frac{\ud
      D}{\ud z_1}\ud z_1},
\end{equation}
where  $D$ is  the comoving  distance of  Sample 1  galaxy,  $r_p$ the
projected  physical separation, $z_l$  and $z_u$  the lower  and upper
limits  of the redshift  interval in  consideration.  If  $\ud N_1/\ud
z_1$ and $\ud D/\ud z_1$ change sufficiently slowly with redshift when
compared to $\xi(r_{1,k})$ as a function of $r_{1,k}$, where $r_{1,k}$
depends on  $z_1$, we can take  $\ud N_1/\ud z_1$ and  $\ud D/\ud z_1$
out of  the integral. Thus for  a thin redshift bin  the ratio between
$w(\theta)$ and $w_p(r_p)$ is simply approximated by
\begin{equation}
  \frac{w(\theta)}{w_p(r_p)}  = \left.\frac{\ud  N_1/\ud  z_1}{N_1 \ud
    D/\ud z_1}\right|_{z_1=z_{med}},
\end{equation}
where $r_p=2D  \sin(\frac{\theta}{2})$ and $z_{med}$ is  the median of
the redshift range.  In practice we instead calculate the ratio in the
following  way to take  into account  the redshift  dependence (though
very weak) of $dN_1/dz_1$ and $dN_2/dz_1$:
\begin{equation}\label{eqn:ratio_ana}
\frac{w(\theta)}{w_p(r_p)}=\frac                  {\int_{z_{l}}^{z_{u}}
  {\left(\left.\frac{\ud      N_1/\ud      z_1}{N_1     \ud      D/\ud
      z_1}\right|_{z_1=z}\right)\left(\ud  N_2/\ud  z\right)  \ud  z}}
     {\int_{z_{l}}^{z_{u}} {\left(\ud N_2/\ud z\right) \ud z}}.
\end{equation}

We also require the angular separation $\theta$ to change accordingly
with fixed $r_p$, and thus in this way the angular separation of our
measured $w(\theta)$ changes with the redshift of spectroscopic
galaxies when counting galaxy-galaxy pairs. To be more specific, the
$r_p$ considered here range from $r_p\sim0.1Mpc/h$ to $r_p\sim25Mpc/h$
for LRGs, and from $r_p\sim0.1Mpc/h$ to $r_p\sim20Mpc/h$ for galaxies
in the low-redshift sample, with 13 and 12 intervals of equal size in
logarithmic space. Quantities in the right side of
Eqn.~(\ref{eqn:ratio_ana}) are determined either from data catalogue
directly ($N_2$ and $\ud N_2/\ud z$) or from the luminosity function
analytically ($\ud N_1/\ud z$).

In  order  to  understand  to what  extent  Eqn.~(\ref{eqn:ratio_ana})
holds, we have  performed two tests. In the first  test, we calculate a
linear  power  spectrum  $P_l(k)$  using  the  CMBFAST  code  for  the
cosmology       adopted       here       \citep[]{1996ApJ...469..437S,
  1998ApJ...494..491Z,  2000ApJS..129..431Z}, from which  we calculate
the     nonlinear     power     spectrum     $P_{nl}(k)$     following
\cite{1996MNRAS.280L..19P}.    The  real-space   correlation  function
$\xi(r)$  is then  obtained by  Fourier transforming  $P_{nl}(k)$. The
amplitude of $\xi(r)$ is arbitrarily  given which has no effect on the
ratio of $w(\theta)$ and  $w_p(r_p)$.  Next, the redshift distribution
$\ud N_1/\ud z_1$  for Sample 1 galaxies in  a certain magnitude range
is calculated analytically from the luminosity function of B03.  Using
Eqn.~(\ref{eqn:simplified_wtheta_xir2})   and~(\ref{eqn:wrp_xir})   we
determine $w(\theta)$ and $w_p(r_p)$  for this magnitude range, giving
rise  to the  {\em  true} value  of  their ratio  which  we denote  as
$ratio_{true}$.     When    integrating     the    right    part    of
Eqn.~(\ref{eqn:simplified_wtheta_xir2})  we  fix  $r_p$  and  let  the
binning  of   $\theta$  vary   accordingly.   We  also   calculate  an
approximated      value     for      the     same      ratio     using
Eqn.~(\ref{eqn:ratio_ana}),  which we  denote  as $ratio_{analy}$  and
compare  to  the  true  ratio   in  order  to  test  the  validity  of
Eqn.~(\ref{eqn:ratio_ana}).

\begin{figure}
\label{fig:ratio}\epsscale{0.9}\plotone{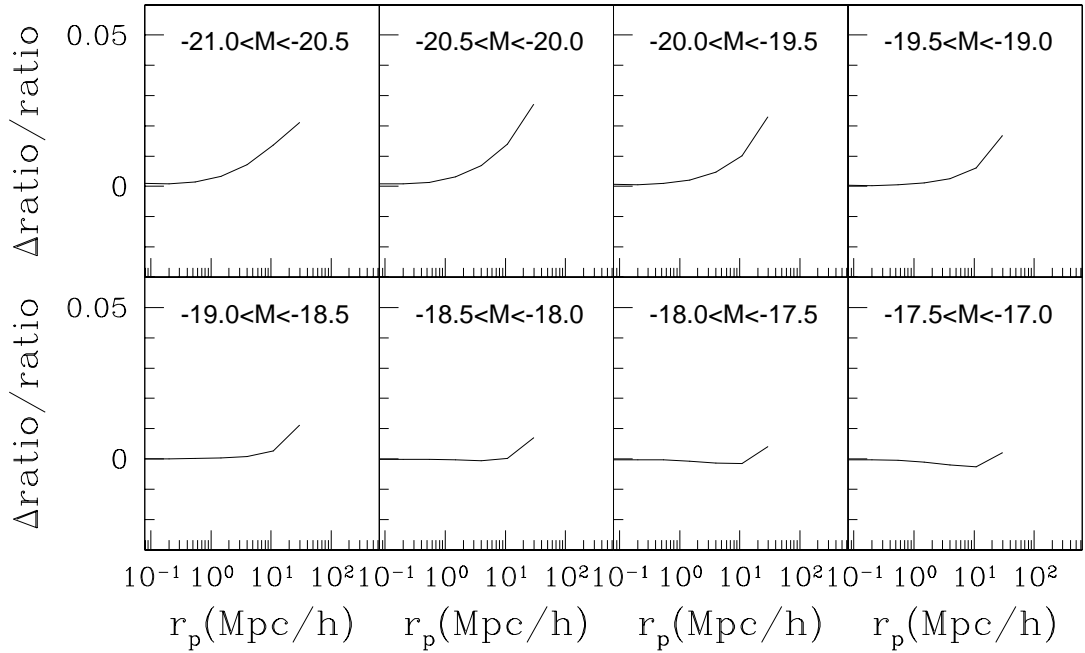}
\caption{Relative difference in $w(\theta)/w_p(r_p)$ ratio
  between the value from Eqn.~(\ref{eqn:ratio_ana}) and the value
  from theoretical calculation (see \S~3.3.1 for details), for 
  spectroscopic galaxies at $0.07<z<0.078$ and photometric galaxies
  at different luminosities, as indicated in each panel.}
\end{figure}

Figure~\ref{fig:ratio}  shows  the  relative  difference  between  the
approximated and  the true  values of the  $w(\theta)/w_p(r_p)$ ratio,
$(ratio_{analy}-ratio_{true})/ratio_{true}$,  for  Sample  2  galaxies
with $0.07<z_2<0.078$  and $-23.0<M_{2}<-21.0$, and  Sample 1 galaxies
in  several  absolute  magnitude   intervals  (as  indicated  in  each
panel). The approximated ratio agrees  quite well with the true value,
at 1\% accuracy or better, for separations $r_p\la$ 10 Mpc and for all
luminosities   considered.   The   discrepancy  increases   at  larger
separations, but well below 3\% level even at the largest scale probed
($\sim30$Mpc). This discrepancy mainly comes from the distant-observer
approximation adopted here. The accuracy of Eqn.~(\ref{eqn:ratio_ana})
is   expected  to   be   better  for   higher   redshifts  where   the
distant-observer  approximation works better.   We thus  conclude that
our approximation in Eqn.~(\ref{eqn:ratio_ana}) works at substantially
high accuracies for our purpose.

A second test that we have done is to apply our method to
spectroscopic samples.  Simply speaking, $w_p(r_p)$ for a
spectroscopic sample can be measured with the redshift information. It
can also be estimated with our method without using the redshift
information for the photometric sample. By comparing the two
$w_p(r_p)$ estimates we are able to understand how well our method
works. For simplicity we consider here a specific case in which a
spectroscopic sample of given luminosity and redshift ranges is
cross-correlated with spectroscopic (for the true $w_p(r_p)$) or
``photometric'' (for the $w_p(r_p)$ obtained by our method) galaxies
in the same ranges. Thus the $w_p(r_p)$ are reduced to
auto-correlation functions.

\begin{figure}
\label{fig:spectests}\epsscale{0.9}\plotone{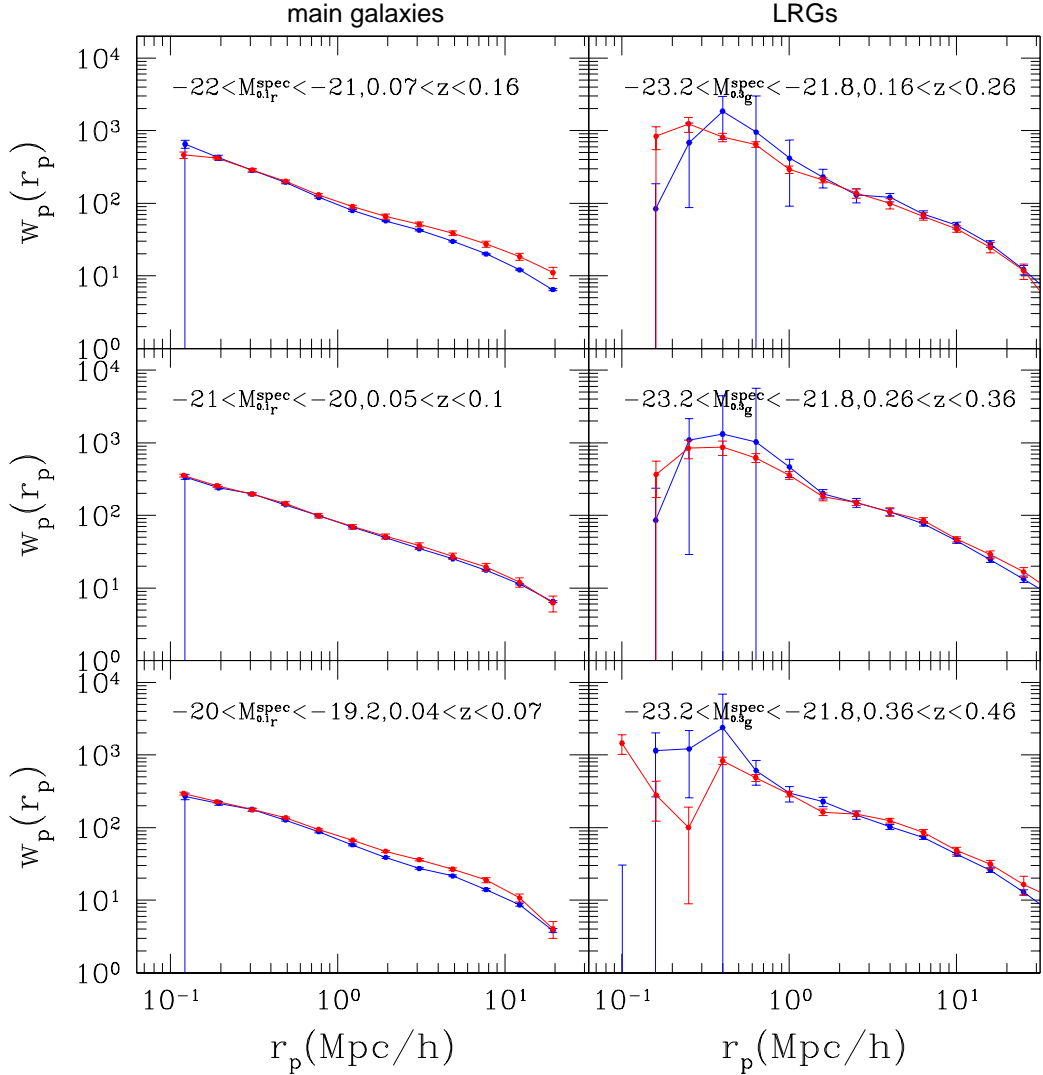}
\caption{Plotted in each panel (red symbols connected by a red line)
is $w_p(r_p)$ of spectroscopic galaxies at 
given luminosity and redshift ranges (as indicated), estimated from 
cross-correlation with photometric galaxies in the same luminosity and
redshift ranges. This is compared to the projected auto-correlation 
correlation function of the same set of spectroscopic galaxies, as
plotted in blue symbols/lines. Panels in the left-hand columns are for
galaxies in the low-redshift sample and those in the right are for LRGs.
}
\end{figure}

The result of this test is shown in Figure~\ref{fig:spectests}, where
we plot the true $w_p(r_p)$ in blue curves and the approximated one in
red for SDSS Main galaxies (left column) and LRGs (right column) at
different luminosities and redshifts (indicated in each panel).  For
LRGs we see good agreement between the two measurements on all scales
and at all redshifts probed (with the difference $<20\%$). A similar
agreement is seen for the low-redshift samples of
$-20<M_{^{0.1}r}<-19$ and $-21<M_{^{0.1}r}<-20$. All these results are
very encouraging.

However, there is large difference ($\sim 50\%$)between the results of the two
methods on large scales ($>5Mpc$/h) for the brightest
($-22<M_{^{0.1}r}<-21$) low reshift sample. The deviation 
may be caused by a coincident correlation between foreground galaxies in the photometric sample and the spectroscopic sample. 
We have performed a further analysis by estimating the
cross-correlation with the foreground, the background, and the right
redshift interval separately, for three low-redshift spectroscopic
samples (corresponding to the left-hand panels in
Fig.~\ref{fig:spectests}). We find that the contamination comes mainly
from the foreground for the brightest sample ($-22<M_{^{0.1}r}<-21$),
and from the background for the faintest sample
($-20<M_{^{0.1}r}<-19$).  For both samples, the projected
cross-correlation $w_p(r_p)$ with the foreground (for the brightest
sample) or the background (for the faintest sample) shows weak
dependence on scale.  When compared to the true $w_p(r_p)$, the
cross-correlation with the foreground/background is negligible on
small scales, $\sim50\%$ smaller at $\sim10$Mpc$/h$ and compatible at
$\sim20$Mpc$/h$. This result clearly shows that the clustering pattern
of the forground/background can contaminate the angular cross
correlation function stochastically. This also explains why we can
measure the projected correlation function for the LRG sample accurately, because
the foreground/background galaxies are in big cosmic volumes and thus
have weak correlations themselves.

We conclude that the accuracy of our method relies on the key
assumption that foreground/background galaxies have weak correlation
with the spectroscopic galaxies. This assumption is valid for many
real observations, especially for spectroscopic samples at
intermediate or high redshift. This is why we can recover the project
correlation function for LRG samples on all scales. For the low
redshift samples, our method works for the sample of luminosity $M_*$,
since the forground/background galaxies are relatively small in number
compared with those at the redshift of the spectroscopic sample. For
the bright low redshift sample, the foreground galaxies are located in
a small volume and are more numerous, and their clustering pantern can
bias the estimation of the projected function. This effect is smaller
for small scales, which is the reason why we can measure the projected
function accurately on scales smaller than $\sim1$ Mpc$/h$.

\subsubsection{Indirect conversion through $w_p(r_p)n$}

We propose a second method here for estimating $w_p(r_p)$. Rather than
directly converting  $w(\theta)$ to  $w_p(r_p)$, we first  convert the
former  to a projected  density profile  $w_p(r_p)n_1$, from  which we
then  estimate  $w_p(r_p)$  by  calculating analytically  the  spatial
number density  $n_1$ from the luminosity  function.  $w_p(r_p)n_1$ is
obtained as a byproduct without suffering from uncertainties in galaxy
luminosity function.

In this method we estimate a {\em weighted} angular correlation
function $w(\theta)_{weight}$ instead of the traditional function
$w(\theta)$ as discussed above.  This is measured using the same
estimators given in Eqn.~(\ref{eqn:standard_estimator})
and~(\ref{eqn:hamilton_estimator}), except that each spectroscopic
galaxy in Sample 2 is weighted by $D^{-2}$, inverse of the square of
comoving distance for Sample 2 galaxies.  It can be easily proved with
Eqn.~(~\ref{eqn:ratio_ana}) that for a thin redshift interval of the
spectroscopic sample (Sample 2), $N_1 w(\theta)_{weight}/\Omega$
equals to the projected density profile $w_p(r_p)n_1$, i.e.,
\begin{equation}\label{eqn:wrpn_wtheta}
w_p(r_p) n_1=\frac{N_1 w(\theta)_{weight}}{\Omega},
\end{equation}
where $N_1$ and $n_1$ are the number and number density of photometric
galaxies in Sample 1, and $\Omega$  the total sky coverage of Sample 1.
Given  the  projected   density  profile  $w_p(r_p)n_1$  estimated  by
Eqn.~(\ref{eqn:wrpn_wtheta}) as well as  the spatial number density of
galaxies $n_1$  analytically calculated from  the luminosity function,
we finally  estimate the projected correlation  function $w_p(r_p)$ by
dividing  $w_p(r_p)n_1$ by  $n_1$.   We emphasize  here the  projected
density profile  $w_p(r_p)n_1$ does  not suffer from  uncertainties in
luminosity  function, because  all  quantities in  the  right side  of
Eqn.~(\ref{eqn:wrpn_wtheta}) can be obtained from data.

To calculate  $n_1$, we adopt  the luminosity evolution model  and the
luminosity  function at $z=0.1$  from B03  when doing  calculation for
spectroscopic  galaxies at  low redshifts  (samples selected  from the
SDSS  Main  galaxy   catalogue  as  listed  in  Table~\ref{tab:lowz}).
Considering that the  evolution model of B03 is  based on low-redshift
data ($z\la 0.25$), which might  not be suitable for higher redshifts,
we adopt the evolution model of \cite{2007ApJ...665..265F} (here after F07)
for our LRG samples.  Moreover, we
need to convert the F07 model from $B$-band to the ${^{0.1}r}$-band at
which  our galaxies  are observed.   Assuming  that the  slope of  the
luminosity  evolution  doesn't  depend  on  waveband,  we  obtain  the
$^{0.1}r$-band  luminosity  evolution  model  by simply  shifting  the
amplitude of the $B$-band model from F07 so as to have an amplitude at
$z=0.1$ which is equal  to the amplitude of the  $^{0.1}r$-band model of
B03.  In this  manner the slopes of the  two models remain unmodified,
which are $Q=-1.23$ for F07  and -1.62 for B03 respectively (see their
papers for details)\footnote{We have repeated our analysis for LRGs,
adopting $Q=-1.62$ instead of $Q=-1.23$, and obtained similar results. 
}.

In conclusion, the projected cross-correlation function $w_p(r_p)$ can
be   measured  either   from   Eqn.~(\ref{eqn:ratio_ana})  by   direct
conversion  of $w(\theta)$,  or  from Eqn.~(\ref{eqn:wrpn_wtheta})  by
indirect conversion through  estimating $w_p(r_p)n$. After having made
extensive  comparisons, we  found that  the two  methods give  rise to
almost identical results. In what  follows we choose to use the second
method  only,  as  it  simultaneously  provides  both  $w_p(r_p)$  and
$w_p(r_p)n$.

\subsection{Division and combination of redshift subsamples}

In this subsection we address an important issue  which we have ignored
so  far.  As  mentioned above,  our method  for  selecting photometric
galaxies according to  luminosity and redshift is valid  only when the
redshift interval  $z_2-z_1$ is  small enough.  However,  the redshift
intervals used  to select our spectroscopic  galaxy samples apparently
do not satisfy  this condition.  For example, for  a redshift range of
$0.16<z<0.26$    and    an     absolute    magnitude    interval    of
$M_{p,2}-M_{p,1}=0.5$, the photometric  galaxies selected will cover a
much  broader apparent  magnitude range,  $m_{p,2}-m_{p,1}=1.7$.  

Our  solution   is  to  further   divide  the  galaxies  in   a  given
spectroscopic  sample into a  number of  subsamples which  are equally
spaced  in  redshift  (hereafter  called  redshift  sub-shells).   See
Tables~\ref{tab:lrg}  and ~\ref{tab:lowz} for  the number  of redshift
sub-shells adopted  for our samples.   For a given sample,  we measure
the  weighted angular cross-correlation  function $w(\theta)_{weight}$
(see above)  for each  sub-shell separately by  cross-correlating with
galaxies selected from the  photometric catalogue in the way described
above  according to  the expected  luminosity range  and  the redshift
range of the sub-shell.  Each $w(\theta)_{weight}$ measurement is then
converted  to   give  the  corresponding   projected  density  profile
$w_p(r_p)n$  as  well  as  the  projected  cross-correlation  function
$w_p(r_p)$,  using the  second method  described above.   Estimates of
these  quantities for  the sub-shells  are then  averaged to  give the
estimates for their parent sample  as a whole.  In this procedure each
sub-shell  is  weighted  by  $V_i/\sigma_i^2$, with  $V_i$  being  the
comoving  volume covered  by the  $i$th sub-shell  and  $\sigma_i^2$ the
variance of $w_p(r_p)n$  or $w_p(r_p)$ of the sub-shell.   In order to
estimate $\sigma_i$  we have generated 100 bootstrap  samples for each
sub-shell. The variance $\sigma_i$ of a sub-shell is then estimated by
the  $1\sigma$  scatter  between  all its  bootstrap  samples.   This
weighting scheme ensures the  averaged $w_p(r_p)n$ or $w_p(r_p)$ to be
determined largely by sub-shells with relatively large volume and high
signal-to-noise  ratio (S/N)  measurements, thus  effectively reducing
the overall sampling noise and cosmic variance.

 In order  to increase  the accuracy  of our method,  one may  want to
 increase the number of sub-shells  for a given redshift range, at the
 cost  of  increasing both  the  sampling  noise  and the  large-scale
 structure noise (the {\em cosmic  variance}). In practice, we split a
 spectroscopic   sample   into   redshift  sub-shells   by   requiring
 $\Delta{z}/{z}\lesssim  0.1$, where $\Delta{z}$  is the  thickness of
 the sub-shells and $z$ is the mean redshift of the sample.  With this
 restriction  the difference  between $m_2-m_1$  and  $M_2-M_1$ ranges
 from $\sim0.25$ occurring for low-redshift sub-shells to $\sim0.1$ for
 high-redshift ones.  Extensive tests show that our results are robust
 to reasonable  change of the thickness of  sub-shells.  For instance,
 taking {\tt Sample L1}  from Table~\ref{tab:lrg} as the spectroscopic
 sample,  the  cross-correlation function  measured  by 10  sub-shells
 differs from the one of 5  sub-shells by at most 10\% for photometric
 galaxies with $-22<M<-21.5$,  and by only about 3\%  for the faintest
 luminosity bin ($-19.5<M<-19.0$).

\subsection{Error estimation}

We estimate the error in the averaged $w_p(r_p)n$ or $w_p(r_p)$ by
\begin{equation}\label{eqn:error}
\Delta=\left\{
\begin{array}{rl}
\sqrt{\sigma^2\times   \frac{{\chi}^2}{\mathtt{dof.}}}   &   \text{if}
\frac{{\chi}^2}{\mathtt{dof.}}>1\\       \sigma       &      \text{if}
\frac{{\chi}^2}{\mathtt{dof.}}\leq 1,
\end{array} \right.
\end{equation}

with
\begin{eqnarray}
  \sigma^2                   &                   =                   &
  \sum\limits_{i=1}^{N_{sub}}{\frac{V_i^2}{\sigma_i^2}}\left/\left(\sum\limits_{i=1}^{N_{sub}}{\frac{V_i}{\sigma_i^2}}\right)^2\right.,\label{eqn:error2}
  \\             \chi^2/dof.             &             =             &
  \frac{1}{N_{sub}-1}\sum\limits_{i=1}^{N_{sub}}\left(x_i-x_{avg}\right)^2\sigma_i^{-2},\label{eqn:error3}
\end{eqnarray}
where the  sum goes over all  the sub-shells of  a given spectroscopic
sample; $x_i$ is  the measurement of $w_p(r_p)n$ or  $w_p(r_p)$ of the
$i$th  sub-shell and  $x_{avg}$ the  average measurement  for  all the
sub-shells  as  a  whole;  $N_{sub}$  is  the  number  of  sub-shells.
Overall,  Eqn.~(\ref{eqn:error}) should  be able  to include  both the
volume effect  (through factor $V_i$) and the  sampling noise (through
$\sigma_i$), thus providing a reasonable estimate of the errors in our
measurements.        By       weighting       the       error       by
$\sqrt{\frac{{\chi}^2}{\mathtt{dof.}}}$, we mean  to take into account
the large variation from sub-shell to sub-shell in some cases.

\begin{figure}
\label{fig:sub_shells}\epsscale{0.9}\plotone{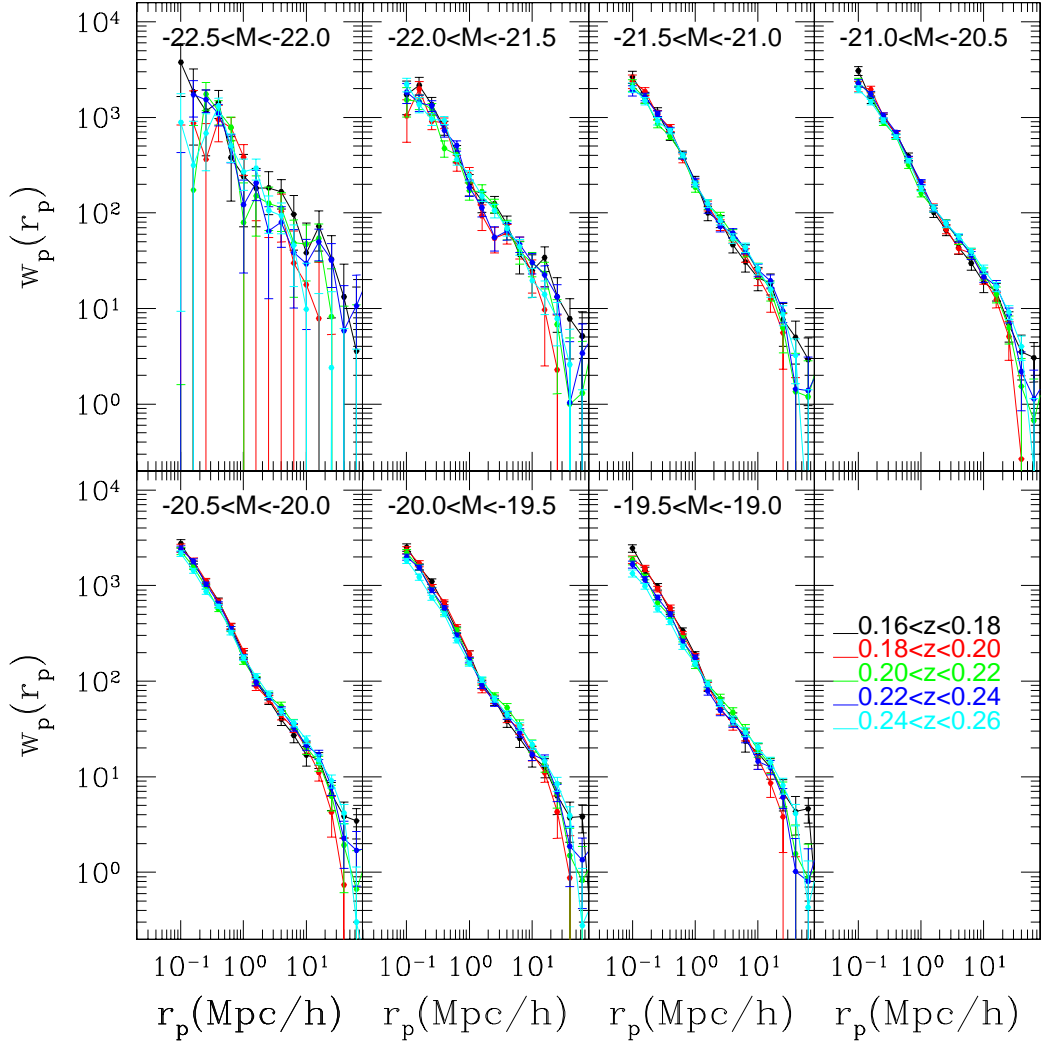}
\caption{Projected cross-correlation  function $w_p(r_p)$ for redshift
  sub-shells of Sample L1 in Table~\ref{tab:lrg}, estimated by
  cross correlating  each  sub-shell  with  photometric  galaxies  in
  different luminosity intervals as indicated in each panel. Different
  lines are for different sub-shells  with their redshift ranges 
  indicated in the bottom-right panel.}
\end{figure}

To better  understand the  error contribution from  different redshift
sub-shells,  we  plot  in Figure~\ref{fig:sub_shells}  the  $w_p(r_p)$
measurements  for  {\tt  Sample  L1}  listed  in  Table~\ref{tab:lrg}.
Different  panels  correspond to  photometric  galaxies in  different
luminosity intervals.  In  each panel, we plot $w_p(r_p)$  for all the
five  sub-shells   with  their   redshift  ranges  indicated   in  the
bottom-right panel.  Error bars on the $w_p(r_p)$ curves are estimated
using  the   bootstrap  resampling  technique,   i.e.   $\sigma_i$  in
Eqn.~(\ref{eqn:error2})  and~(\ref{eqn:error3}).   We  see  that,  for
photometric galaxies  at fixed luminosity,  $w_p(r_p)$ measurements of
different sub-shells are almost on  top of each other, indicating that
the scatter between sub-shells is  fairly small. This again shows that
the correlation functions measured  with our method are insensitive to
the number of redshift sub-shells.

We note  that the overall  error increases rapidly with  luminosity at
the bright end. This reflects not only the sampling noise of the small
samples, but  more importantly,  also an effect  of a  huge foreground
population   which   significantly   suppresses  the   {\em   angular}
cross-correlation signals.  Letting  $w^\prime(\theta)$ be the angular
correlation function  between a spectroscopic  sample with $z_1<z<z_2$
and  a photometric  sample including  galaxies of  all  redshifts, and
$w(\theta)$ the one between the same spectroscopic sample and a sample
of photometric  galaxies within $z_1<z<z_2$, one can
easily show that
\begin{equation}
  w'(\theta)=\frac{N_{GS}}{N_G}w(\theta),
\end{equation}
where $N_G$  and $N_{GS}$ are  respectively the number  of photometric
galaxies in the full sample  and in the redshift range $z_1<z<z_2$. In
this case the  estimated correlation signal is suppressed  by a factor
of  $\frac{N_{GS}}{N_G}$, a  large effect  in particular  when $N_G\gg
N_{GS}$ as in bright samples.

\begin{figure}
\label{fig:dndz_z}\epsscale{0.9}\plotone{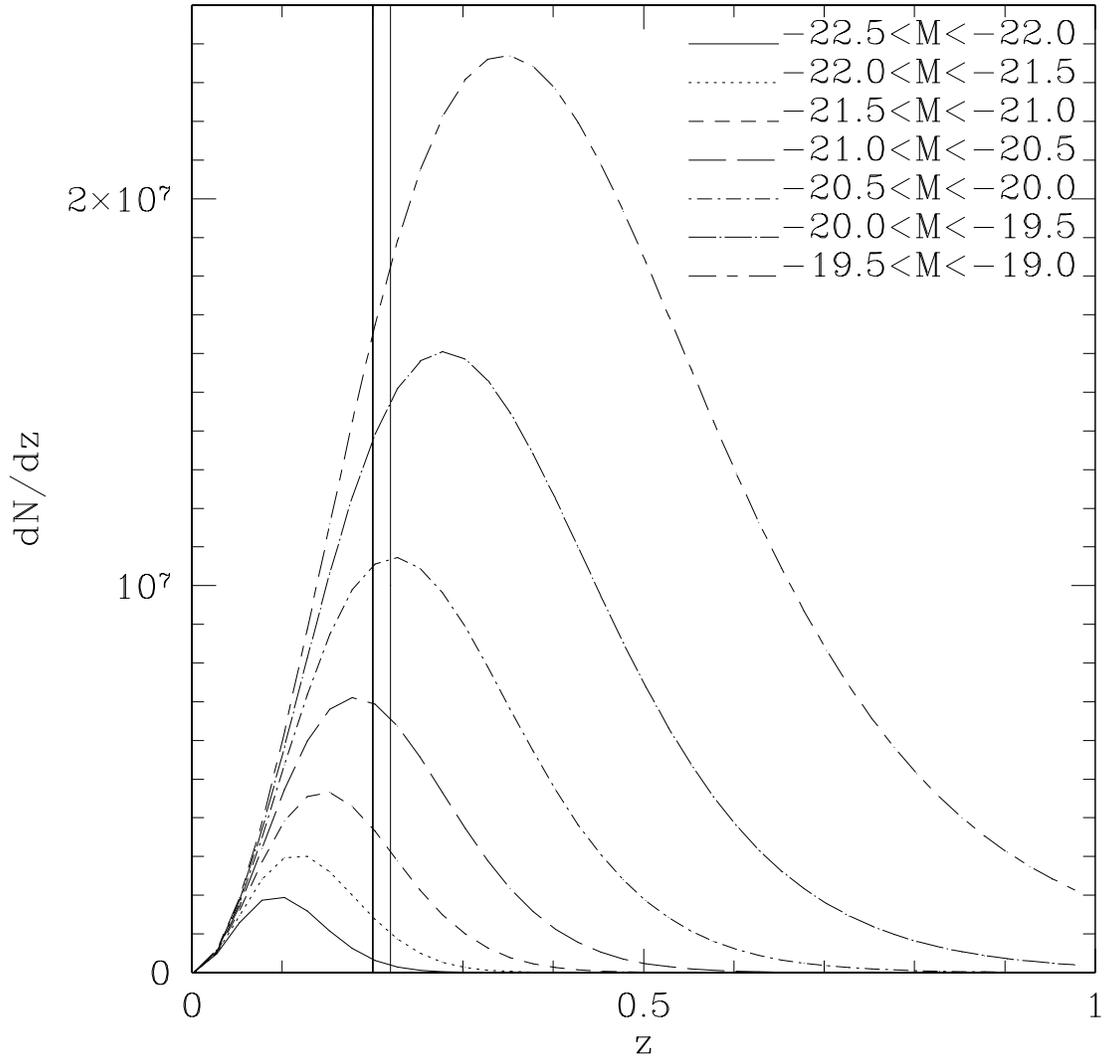}
\caption{Redshift distribution as predicted by the luminosity function 
  from \cite{2003ApJ...592..819B} for photometric galaxies that are 
  expected to fall in  the
  indicated luminosity intervals if they were located in the redshift
  range $0.2<z<0.22$. The two vertical lines mark the redshift range.}
\end{figure}

This  is explained  more clearly  in Figure~\ref{fig:dndz_z}  where we
plot  the redshift  distribution  as calculated  using the  luminosity
function of B03 for  photometric galaxies which are selected according
to   luminosity   and  redshift   using   the   method  described   in
\S~\ref{sec:selecting_sat}.   Plotted  in   different  lines  are  the
distributions  for  different luminosity  ranges  with redshift  range
fixed  to   $0.2<z<0.22$.   According   to  our  method   of  dividing
photometric  sample into  luminosity subsamples,  photometric galaxies
selected to  serve our  purpose of a  certain luminosity bin  have the
desired luminosity  at the  chosen redshift. As  can be seen  from the
figure,      in       the      brightest      luminosity      interval
($-22.5<M_{^{0.1}r}<-22.0$, solid  line) only a small  fraction of the
galaxies  are located within  the expected  redshift range,  where the
majority of galaxies are  intrinsically fainter. It is this population
that suppresses the angular  cross-correlations that we measure at the
bright end, leading  to large uncertainties in $w_p(r_p)$,  as seen in
Figure~\ref{fig:sub_shells}.

Before we apply our method to SDSS data in the next section, we should
point  out  that,  although  we  have carefully  considered  both  the
sampling  noise  and  the   large-scale  structure  noise,  our  error
estimation doesn't include the  projection effect caused by background
and foreground  galaxies.  Thus when interpreting  our $w_p(r_p)n$ and
$w_p(r_p)$ presented below, one should  keep in mind that their errors
are underestimated  to varying degrees, depending on  the redshift and
luminosity we consider.

\section{Applications to SDSS galaxies}
\label{sec:apply to sdss}

\subsection{Projected cross-correlations and density profiles of LRGs}

\begin{figure}
\label{fig:profile_lrg}\epsscale{0.9}\plotone{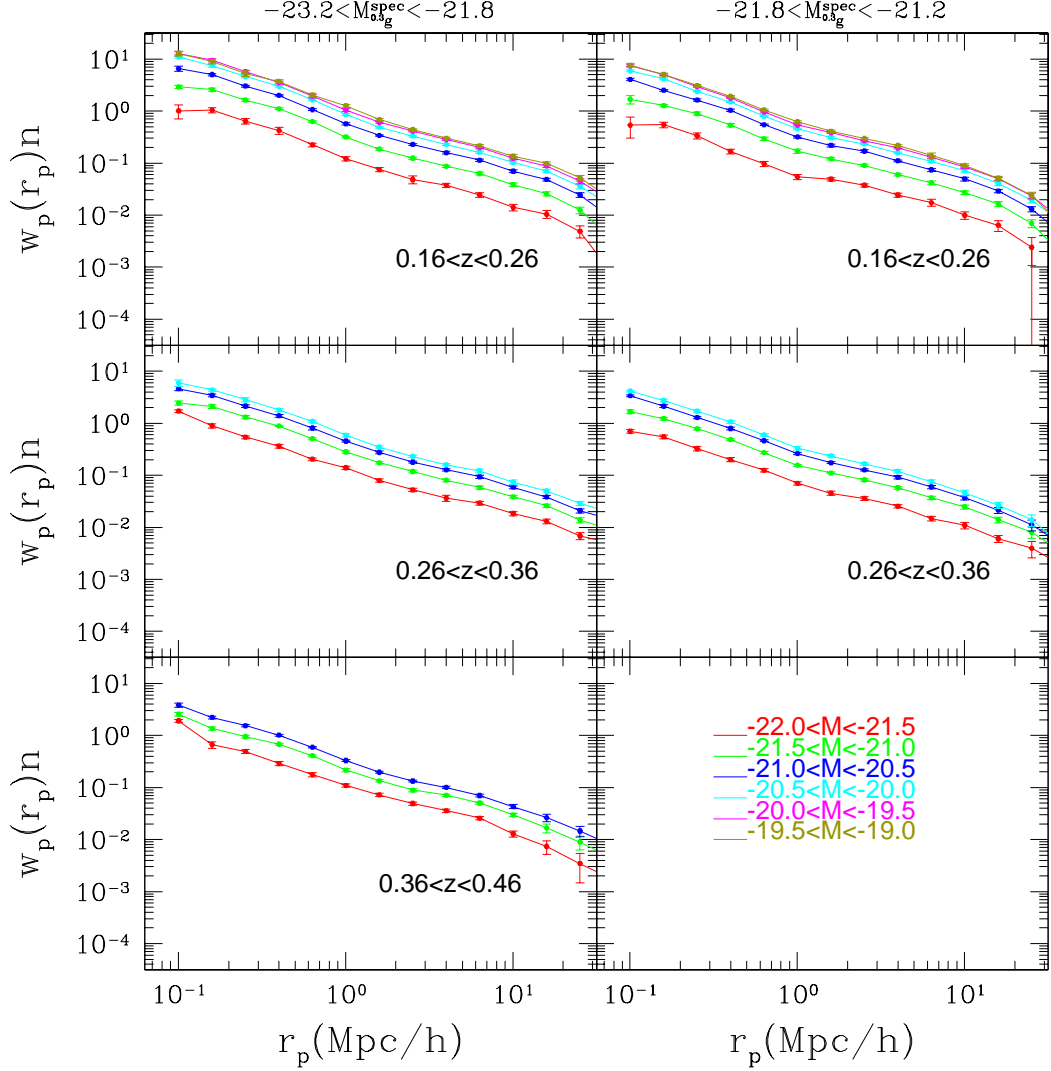}
\caption{Projected   density   profile   $w_p(r_p)n$   in   units   of
  $Mpc^{-2}h^2$   surrounding   LRGs   with   different   luminosities
  (indicated  above  the  figure)  and redshifts  (indicated  in  each
  panel), as  traced by galaxies  of different luminosities  (shown in
  different lines in each panel and indicated in the bottom-right panel).}
\end{figure}

In Figure~\ref{fig:profile_lrg} we  show the projected density profile
$w_p(r_p)n$  for  LRGs  in   different  intervals  of  luminosity  and
redshift, as  traced by surrounding photometric  galaxies of different
luminosities.      Results    are     plotted     for    LRGs     with
$-23.2<M_{^{0.3}g}<-21.8$  in panels on  the left  and for  those with
$-21.8<M_{^{0.3}g}<-21.2$ on the right, with panels from top to bottom
corresponding to  different redshift bins.  The  $w_p(r_p)n$ traced by
photometric galaxies  in different  luminosity ranges are  shown using
different lines  as indicated  in the bottom-right  panel.  As  can be
seen,  the projected number  density of  galaxies around  central LRGs
decreases  as  their  luminosity  increases.   This is  true  for  all
redshifts  and  all scales  probed.   In  particular, such  luminosity
dependence  is  weak  for  galaxies fainter  than  the  characteristic
luminosity of the  Schechter luminosity function ($M_{^{0.1}r}=-20.44$
for  the SDSS),  and becomes  remarkable for  brighter  galaxies.  The
change of the amplitude mainly reflects the number density of galaxies
as a  function of  their luminosity.  From  the figure one  can easily
read out  the number  of galaxies in  the host  halo of the  LRGs. For
example, on average there are about 3 galaxies of $-20.5<M<-20$ in the
host halo of  a LRG in {\tt Sample L1}, assuming  the host halo radius
is $0.3h^{-1}$Mpc.

\begin{figure}
\label{fig:wrp_lrg}\epsscale{0.9}\plotone{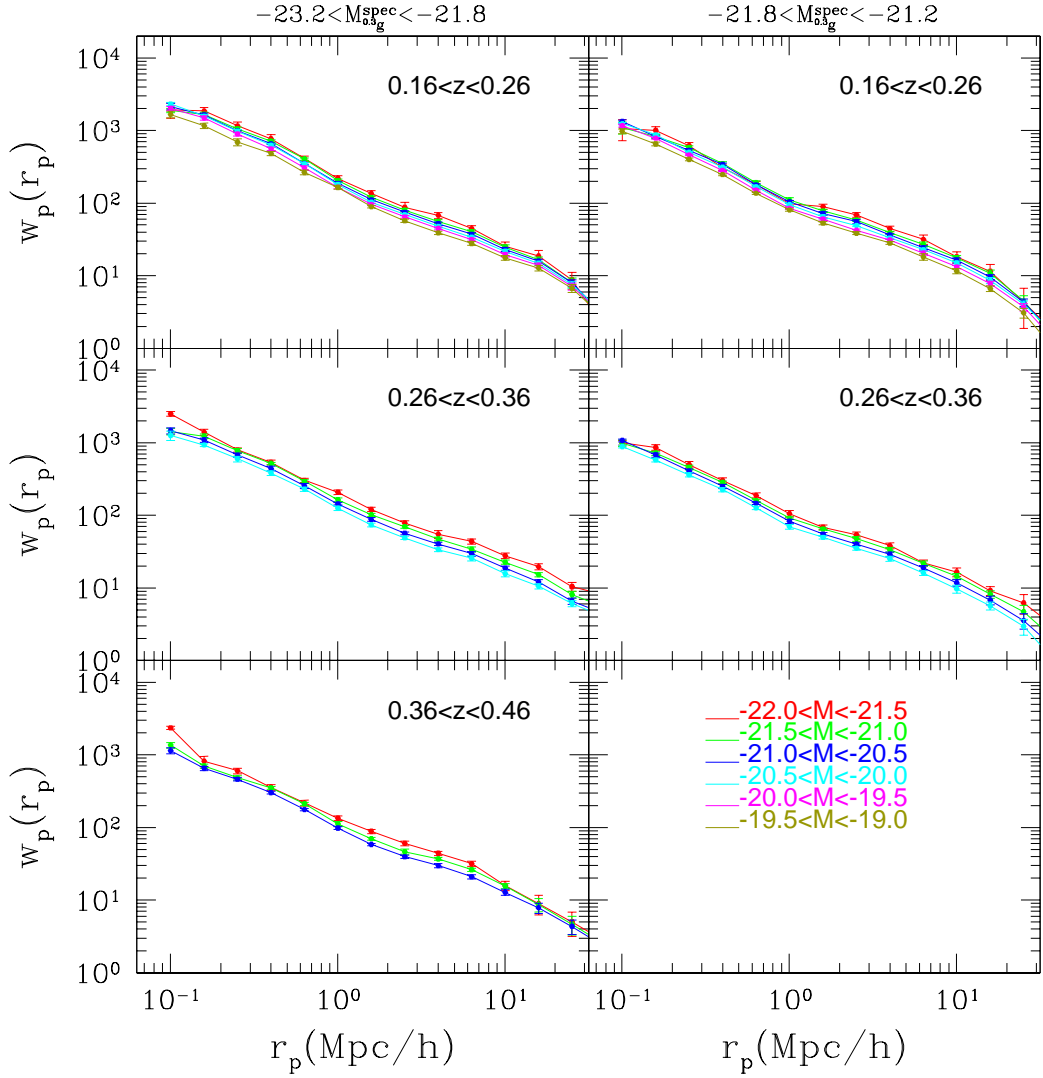}
\caption{Projected cross-correlation  function $w_p(r_p)$ measured for
  the same  set of LRG samples  and the same  intervals of photometric
  galaxy luminosity as in the previous figure.}
\end{figure}

In   Figure~\ref{fig:wrp_lrg}  we  show   projected  cross-correlation
function $w_p(r_p)$  obtained from the  $w_p(r_p)n$ measurements shown
in Figure~\ref{fig:profile_lrg},  for the same set of  LRG samples and
the same  intervals of photometric  galaxy luminosity. At  fixed scale
and redshift,  the amplitude  of $w_p(r_p)$ increases  with increasing
luminosity,  a  trend  which  has  already been  well  established  by
previous studies.   It is interesting  to see from both  figures that,
although both  $w_p(r_p)n$ and $w_p(r_p)$ show  systematic trends with
luminosity  in amplitude,  their  slope remains  fairly universal  for
given  redshift  and  central  galaxy luminosity,  regardless  of  the
luminosity of surrounding galaxies  we consider.  This provides direct
observational   evidence   that   satellite  galaxies   of   different
luminosities follow  the distribution of  dark matter in the  same way
within their dark matter halos, an assumption adopted in many previous
studies on HOD modeling of galaxy distribution.

\begin{figure}
\label{fig:evolution_lrg}\epsscale{0.9}\plotone{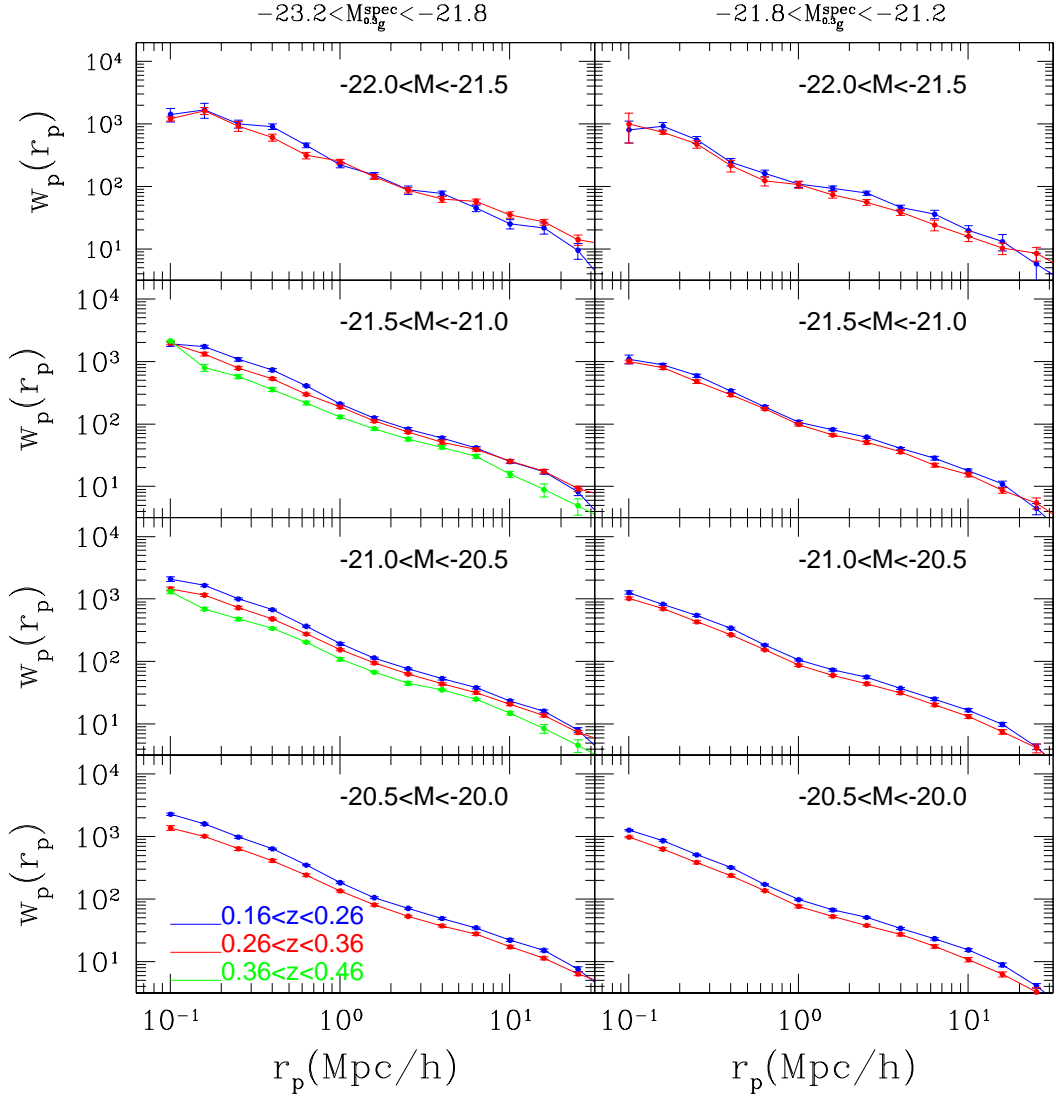}
\caption{Each panel compares $w_p(r_p)$ measured at different
  redshifts (as indicated), but for fixed luminosity ranges for photometric and
  spectroscopic galaxies. The luminosity ranges of spectroscopic
galaxies are indicated above the figure, while those of photometric
galaxies are indicated in each panel.}
\end{figure}

In  Figure~\ref{fig:evolution_lrg}  we plot  the  $w_p(r_p)$ again  in
order  to  explore  the  evolution  with  redshift.   Measurements  of
different  photometric  galaxy  luminosities  are shown  in  different
panels,  while  in  each  panel  we  compare  $w_p(r_p)$  measured  at
different redshifts for fixed luminosity.  Note that in this figure we
have considered  the luminosity  evolution of galaxies,  as we  aim to
study  how the  projected density  distribution  and cross-correlation
around the  LRGs have evolved over  the redshift range  probed.  We do
not               include               $E$-correction              in
Figures~\ref{fig:sub_shells},\ref{fig:profile_lrg}
and~\ref{fig:wrp_lrg}, because  we want the  results there to  be less
affected by  possible uncertainty  in the luminosity  evolution model.
From  Figure~\ref{fig:evolution_lrg}, we  see significant  increase in
the  amplitude  of  $w_p(r_p)$   as  redshift  goes  from  $z=0.4$  to
$0.2$. Taking the $-21.0<M<-20.5$ bin  in the left column for example,
on average  the $w_p(r_p)$  amplitude differs by  a factor of  about 2
between the  result of $z\sim0.2$  (blue curve) and  $z\sim0.4$ (green
curve) at separations $r_p<0.3 h^{-1}$Mpc, the typical boundary of LRG
host halos.

It is  important to understand whether the  significant evolution seen
above  could   be  explained  purely  by  evolution   in  dark  matter
distribution, or  additional processes related  to galaxies themselves
are  necessary.  To  the  end we  have  done a  simple calculation  as
follows. We assume that there  is no merger occurring between galaxies
or between their host halos.  In this case all galaxies and halos have
been evolving in a passive manner,  and the total number of each keeps
unchanged  during  the  period  in consideration.   Thus  galaxies  at
different redshifts  are the same  population which are hosted  by the
same set of halos. We calculate a dark matter density profile averaged
over all  dark matter halos  that are expected  to host LRGs  of given
redshift and luminosity, by
\begin{equation}\label{eqn:halo_profile}
  \rho_{avg}(r)=\frac{\int_{M_{min}}^{M_{max}}{\rho(r,M)  n(M)\ud  M}}
      {\int_{M_{min}}^{M_{max}}{n(M)\ud M}},
\end{equation}
where $n(M)$ is the halo mass function from
\cite{1999MNRAS.308..119S}, and $\rho(r,M)$ is the density profile of
halos of mass $M$ assumed to be in the NFW form
\citep[]{1997ApJ...490..493N}.  We determine the concentration
parameter $c$ of halos following \cite{2009ApJ...707..354Z}. The lower
and upper limits of halo mass ($M_{min}$ and $M_{max}$) are determined
by matching the abundance of LRGs in our sample with that of dark
matter halos given by the Sheth-Tormen mass function
\citep{1999MNRAS.308..119S}.  For this we have adopted the plausible
assumption that the luminosity of a central galaxy is an increasing
function of the mass of its halo.

We  consider  three  redshifts  which  are $z=$0.21,  0.31  and  0.41,
approximately  the median  redshifts  of our  LRG  samples.  The  dark
matter   density   within   $0.3    h^{-1}$   Mpc   around   LRGs   of
$-23.2<M_{^{0.3}g}<-21.8$  is predicted to  increase by  $24.6\%$ from
$z\sim0.4$ to  $\sim0.2$, and $12.1\%$ from  $z\sim0.3$ and $\sim0.2$.
The factor  is $10.6\%$ for  $-21.8<M_{^{0.3}g}<-21.2$ from $z\sim0.3$
to $\sim0.2$.   It is  clear that these  predictions are  much smaller
when compared to what we have  obtained from the SDSS data.  As can be
seen from Figure~\ref{fig:evolution_lrg},  at $r_p<0.3 h^{-1}$Mpc, the
clustering amplitude changes by a factor of $\sim$2 between $z\sim0.4$
and  $\sim0.2$, and  by 20  - 50\%  between $z\sim0.3$  and $\sim0.2$.

This  simple calculation seems to suggest that  the evolution  of galaxy
clustering observed in  this work is caused not  only by the evolution
of the underlying  dark matter, but also by  the evolution of galaxies
themselves. However,  this argument  should not be  overemphasized.
As selected in $r$-band, our photometric sample may be biased to bluer
galaxies as one goes to higher redshifts. This selection effect must
be properly taken into account when one addresses the evolution of
galaxy clustering, but such a detailed modeling is out of the scope of
our current paper.

\begin{figure}
\label{fig:bias_lrg}\epsscale{0.9}\plotone{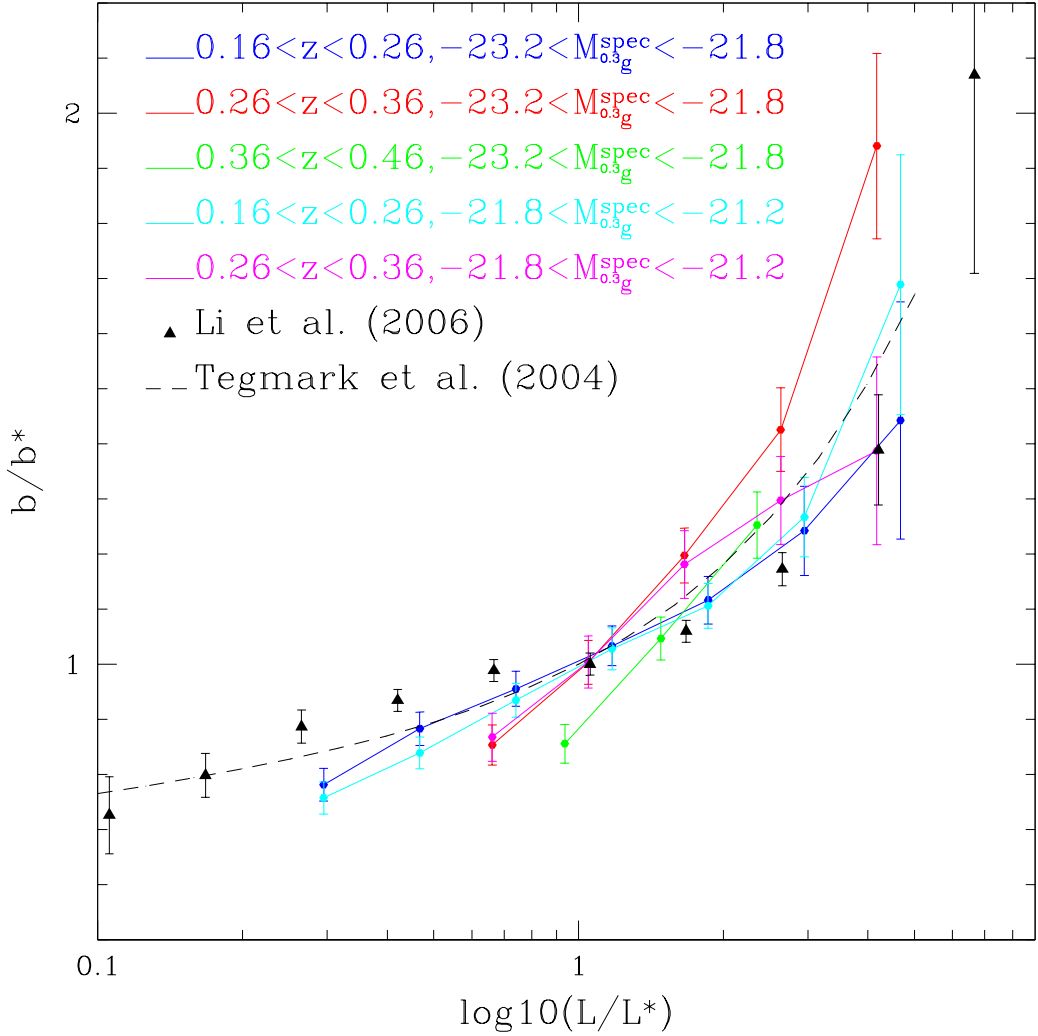}
\caption{Relative bias factor with  respect to $L*$ as a function of
  luminosity.  Bias factors are calculated  from  the
  amplitude  of  $w_p(r_p)$ averaged  over  $2.5h^{-1}Mpc<r<10.0h^{-1}Mpc$,
  and normalized  by $w_p(r_p)$  between the  same LRG  sample  
  and the photometric sample with $-21<M_{^{0.1}r}<-20$.
  Note that for the  $0.36<z<0.46$ bin, the photometric  catalogue  
  becomes  incomplete  when $M>-20.5$,  and so we calculate the relative
  bias  factor  with   respect   to the photometric sample of
  $-21.0<M_{^{0.1}r}<-20.5$.  Different curves are for results obtained 
  with different  LRG subsamples, as indicated. Black triangles and 
  the black dashed line show previous determinations from 
  \cite{2006MNRAS.368...21L} and \citep[]{2004ApJ...606..702T}, which 
  are based on auto-correlation function or power spectrum of the
  SDSS Main galaxy sample with a mean redshift of $z\sim0.1$.}
\end{figure}

Figure~\ref{fig:bias_lrg}   shows   the   relative  bias   factor   of
photometric galaxies  with respect to  $L*$ galaxies as a  function of
luminosity.   Taking a LRG  sample, the  bias factor  for a  sample of
photometric galaxies  at given  luminosity and redshift  is calculated
from  the  amplitude  of  the  $w_p(r_p)$  between  the  LRG  and  the
photometric  samples, normalized  by the  $w_p(r_p)$ of  the  same LRG
sample  with a  photometric sample  selected  by $-21<M_{^{0.1}r}<-20$
($K-$ and $E-$corrected to $z=0.1$)
\footnote{The absolute magnitude range for selecting $L*$ samples varies 
  from sub-shell to sub-shell for the photometric galaxies, so that the 
  corresponding absolute magnitude range at $z=0.1$ is always 
  $-21<M_{^{0.1}r}<-20$.},  and  averaged  over
separations  between $r_p=2.5h^{-1}$Mpc  and $10h^{-1}$Mpc.   The bias
factor so obtained should be  virtually identical to the one estimated
from  the auto-correlation  function of  the same  set  of photometric
galaxies.  In the figure, curves  in different colors refer to results
from  different   LRG  samples  in   Table~\ref{tab:lrg}.   Since  the
photometric sample becomes somewhat  incomplete for $L \lesssim L*$ in
the redshift range $0.36<z<0.46$, the bias factor for this sample (the
green  line   in  the  figure)   is  normalized  with  respect   to  a
$-21.0<M_{^{0.1}r}<-20.5$ sample  instead of the $-21<M_{^{0.1}r}<-20$
one.     Plotted   in    black    triangles   is    the   result    of
\cite{2006MNRAS.368...21L}.  Black dashed line  is a fit obtained from
the SDSS power  spectrum by \cite{2004ApJ...606..702T}.  Relative bias
factors at $0.16<z<0.26$ (Samples L1  and L2) are well consistent with
those  previous studies  at all  luminosities, except  the  bright end
where   our   measurement   is   slightly   lower   than   that   from
\cite{2004ApJ...606..702T}.   Our bias  factor measurements  show that
the luminosity  dependence of galaxy clustering observed  in the local
Universe is very similar to that at intermediate redshift $z\sim0.4$.

\subsection{Projected density profiles and clustering of low-z galaxies}

We have also measured $w_p(r_p)n$  and $w_p(r_p)$ for the eight volume
limited    samples     of    low-redshift    galaxies     listed    in
Table~\ref{tab:lowz} and for different intervals of photometric galaxy
luminosity.  Since  the volume covered by  these spectroscopic samples
is small  due to their  small redshift intervals, the  $w_p(r_p)n$ and
$w_p(r_p)$ measurements  are more noisy  than presented above  for the
LRG samples.  In  order to improve the S/N of  our measurements, for a
given  luminosity  interval of  photometric  galaxies  we combine  the
measurements for  spectroscopic samples  that share a  same luminosity
interval  but   span  different   redshift  ranges.  When   doing  the
combination we  weight each sample  by its comoving volume  divided by
the  variance  of the  measurement,  in the  same  way  as above  when
combining redshift  sub-shells.  The combined  $w_p(r_p)n$ are plotted
in Figure~\ref{fig:profile_lowz}.  We do not include independent plots
for $w_p(r_p)$, which show behaviors very similar to $w_p(r_p)n$.

In Figure~\ref{fig:bias_lowz} we  show the corresponding relative bias
factors,  based on  data  points over  $1.9\le  r_p\le 10  h^{-1}$Mpc.
Results  are plotted  in  blue, red  and  green curves  for the  three
luminosity intervals of spectroscopic galaxies. For comparison we also
repeat     the    bias     factors    from     previous     work    by
\cite{2004ApJ...606..702T}  and  \cite{2006MNRAS.368...21L} which  are
based  on  spectroscopic samples  similar  to  those  used here.   Our
measurements   are   roughly  in   agreement   with  these   previous
determinations.  Our  bias factors from  the three samples  agree with
each  other at  the intermediate  luminosities around  $L_\ast$, while
showing obvious deviations at the  bright and faint ends. Again, these
differences  should  not  be  regarded  as significant  due  to  large
uncertainties.

Similar to what is found for LRGs at $z\sim0.4$, the density profiles
around galaxies in the local universe also shows quite similar slope,
independent of the luminosity of surrounding photometric galaxies.
Unlike in the LRG samples, the spectroscopic galaxies in our
low-redshift samples could be either centrals or satellites of their
host halos.  However, the fraction of satellites should be small as
the spectroscopic objects in the SDSS are relatively bright
\citep[]{2007ApJ...667..760Z}.  Thus our conclusion made above for
$z\sim0.4$ more or less holds for the local Universe, that is,
satellite galaxies of different luminosities are distributed within
their halos in a similar way, if the halos host central galaxies of
similar luminosity.  This is consistent with previous studies
\citep[e.g.][]{2004MNRAS.353..189V, 2006MNRAS.371.1173V,
  2007ApJ...667..760Z} which have revealed a tight relation between
central galaxy luminosity $\langle L_c \rangle$ and halo mass.
Moreover, halo occupation distribution models usually assume galaxy
distribution inside halos to trace their dark matter, based on studies
of satellite distributions in simulations
\citep[e.g.][]{2005ApJ...618..557N, 2006MNRAS.366.1529M}.  Again, our
results provide additional, clear evidence for supporting this
assumption.

\section{Discussion and Summary}

Previous  studies on  galaxy clustering  as a  function  of luminosity
usually make  use of spectroscopic galaxy catalogue,  thus are limited to
relatively bright  galaxies and  low redshifts. In  this work  we have
developed  a new  method which  measures simultaneously  the projected
number density profile $w_p(r_p)n$ and the projected cross-correlation
function $w_p(r_p)$  of a set  of photometric galaxies,  surrounding a
set of spectroscopic  galaxies. We are able to  divide the photometric
galaxies by  luminosity even when  redshift information is
unavailable.  This enables us to measure $w_p(r_p)n$ and $w_p(r_p)$ as
a function of  not only the luminosity of  the spectroscopic galaxies,
but  also  that  of   the  surrounding  photometric  galaxies.   Since
photometric  samples   are  usually  much  larger   and  fainter  than
spectroscopic ones, with our method  one can explore the clustering of
galaxies to fainter luminosities at high redshift.

We have applied  our method to the SDSS data including a sample of  
$10^5$ luminous red galaxy
(LRGs) at $z\sim0.4$ and a sample  of about half a million galaxies at
$z\sim0.1$. Both are cross-correlated with an SDSS photometric
sample consisting of about  20  million  galaxies down  to  $r=21$.  
We  have investigated the dependence of  $w_p(r_p)n$ and $w_p(r_p)$  on galaxy
luminosity   and  redshift,   by  dividing   both   spectroscopic  and
photometric galaxies  into various luminosity  intervals and different
redshift ranges. 

The conclusions of this paper can be summarized as follows.
\begin{itemize}

\item  We develop a new method which measures the projected density
distribution $w_p(r_p)n$ of photometric galaxies surrounding a set of
spectroscopically-identified galaxies, and simultaneously the
projected cross-correlation function $w_p(r_p)$ between the two
populations.  In this method we are able to divide the photometric
galaxies into subsamples in luminosity intervals even when redshift
information is unavailable, enabling us to measure $w_p(r_p)n$ and
$w_p(r_p)$ as a function of not only the luminosity of the
spectroscopic galaxy, but also that of the photometric
galaxy. Extensive tests show that our method can measure $w_p(r_p)$ in
a statistically unbiased way. The accuracy of the measurement depends
on the validity of the assumption inherent to the method that the
foreground/background galaxies are randomly distributed and are thus
uncorrelated with those galaxies of interest. Therefore, our method can be
applied to the cases where foreground/baground galaxies are
distributed in large volumes, which is usually valid in real
observations.

\item We find that, for a spectroscopic sample at given luminosity and
  redshift,  the projected  cross-correlation  function and  projected
  density profile as traced by photometric galaxies show quite similar
  slope  to   each  other,  independent  of  the   luminosity  of  the
  photometric  galaxies.  This  indicates that  satellite  galaxies of
  different luminosities are distributed in a similar way within their
  host  dark  matter  halos.  This  is  true  not  only  for  LRGs  at
  intermediate redshifts  which are  mostly central galaxies  of their
  halos, but also for the  general population of galaxies in the local
  Universe.   Our  result   provides  observational  support  for  the
  assumption commonly-adopted  in halo occupation  distribution models
  that  the   distribution  of   galaxies  follows  the   dark  matter
  distribution within their halos.

\item The relative bias factors are estimated for photometric galaxies
  as a function of luminosity and redshift. In particular, we measured
  the bias factors  of such kind, for the first  time, for galaxies at
  intermediate redshift  ($z\sim0.4$) over a wide  range in luminosity
  ($0.3L_\ast<L<5L_\ast$).

\end{itemize}

There  have been previous  studies of  measuring galaxy  clustering by
cross-correlations  with  imaging   data.  Although  the  methods  and
purposes of  these studies  are different from  ours, it is  worthy of
mentioning   them   and  pointing   out   the   findings  in   common.
\cite{2005ApJ...619..178E} measured  the mean overdensity  around LRGs
in an  earlier SDSS release  by cross-correlating 32,000 LRGs  with 16
million photometric galaxies,  using the deprojecting method developed
in   \citep{2003ApJ...586..718E}.    The   authors   were   aimed   at
understanding the  scale- and luminosity-dependence  of the clustering
of LRGs, thus using only $L_\ast$ galaxies from the photometric sample
as a tracer  of the surrounding distribution. In  our work we consider
luminosity dependence for both LRGs and photometric galaxies, and this
is  why we  have made  considerable efforts  on  selecting photometric
galaxies   of    specific   luminosities.   Our    measurements   from
cross-correlation  with $L_\ast$  galaxies show  strong  dependence in
$w_p(r_p)$  amplitude  on  LRG  luminosity,  as  well  as  an  obvious
transition at around 1 Mpc$/h$, which are in agreement with what those
authors find.
 
\cite{2006ApJ...644...54M}  measured   the  cross-correlation  between
$\sim$25,000 LRGs and  an imaging sample, in order  to correct for the
effect of  fiber collisions on  their small-scale measurements  of LRG
auto-correlations.   They find  that  the real-space  auto-correlation
function of LRGs, $\xi(r)$, is  surprisingly close to a $r^{-2}$ power
law over  more than 4 orders  of magnitude in separation  $r$, down to
$r\sim  15$  kpc$/h$. We  don't  have  a  measurement of auto-correlations
$\xi(r)$  or
$w_p(r_p)$ for LRGs, but our  results are not inconsistent with theirs
in the sense  that the projected cross-correlation of  our LRG samples
continuously increases at  such small scales, with a  slope similar to
that on large  scales. In a recent work, \citet{White-10}
studied the  clustering of massive
galaxies at  $z\sim0.5$ using  the first semester  of BOSS  data.  The
authors computed the projected cross-correlation between their imaging
catalogue  and the  spectroscopic one,  as an  additional  analysis to
emphasize  that there  is  a  significant power  on  scales below  0.3
Mpc$/h$ where their  spectroscopy-based measurements suffer from fiber
collisions. This is  obviously consistent with what we  have seen from
our measurements.

Our method can  be applied and extended to  many important statistical
studies of  galaxy formation and evolution. By  dividing galaxies into
red and blue populations in  the photometric sample according to their
colors, we can quantify the evolution of the blue fraction of galaxies
in clusters and groups from redshift 0.4 to the present day \citep[i.e
  the      Butcher-Oemler     effect,     e.g.][]{2003PASJ...55..739G,
  2004MNRAS.351..125D}.   Wide   deep  photometry  surveys,   such  as
Pan-Starrs        \citep[]{2004SPIE.5489...11K}        and        LSST
\citep[]{2002SPIE.4836...10T},   will  be   available   in  the   next
years. Combining  such surveys with large  spectroscopic samples, such
as  BOSS LRG  samples, will  allow one  to explore  the  clustering of
galaxies from $z=0$ up to $z=1$  for a wide range of luminosities. The
WISE \citep[]{2010AJ....140.1868W}  will produce an  all-sky catalogue
of   infrared  galaxies.    By   combining  this   survey  with   SDSS
spectroscopic  samples,  one  can  study  how  infrared  galaxies  are
distributed relative to the network of optical galaxies.

\acknowledgments

This work is supported by  NSFC (10821302, 10878001), by the Knowledge
Innovation  Program of CAS  (No.  KJCX2-YW-T05),  by 973  Program (No.
2007CB815402), and by  the CAS/SAFEA International Partnership Program
for Creative Research Teams (KJCX2-YW-T23).


\tablenum{1}\label{tab:lrg}
\begin{deluxetable}{lccccc}
\label{tbl:lrg_samples}\tablecolumns{6}\tablewidth{0pc}
\tablecaption{LRG  samples   selected  according  to   luminosity  and
  redshift}  \tablehead{  \colhead{}   &  \colhead{}  &  \colhead{}  &
  \colhead{}      &      \multicolumn{2}{c}{Sub-shell     information}
  \\  \cline{5-6}  \colhead{\sc  Sample} &  \colhead{$M_{^{0.3}g}$}  &
  \colhead{$z$}  &  \colhead{Num.  of  galaxies}  & \colhead{Num.   of
    sub-shells}  &   \colhead{$\Delta{z}_{sub}$}  \\  \colhead{(1)}  &
  \colhead{(2)}  &  \colhead{(3)} &  \colhead{(4)}  & \colhead{(5)}  &
  \colhead{(6)} } \startdata L1 & (-23.2,-21.8) & (0.16,0.26) & 5592 &
5 & 0.02\\ L2 & (-21.8,-21.2) &  (0.16,0.26) & 16832 & 5 & 0.02\\ L3 &
(-23.2,-21.8) & (0.26,0.36) & 11647 &  5 & 0.02\\ L4 & (-21.8,-21.2) &
(0.26,0.36) &  26301 & 5 &  0.02\\ L5 & (-23.2,-21.8)  & (0.36,0.46) &
17502 & 5 & 0.02\\ \enddata
\end{deluxetable}

\tablenum{2}\label{tab:lowz}
\begin{deluxetable}{lccccc}
\label{tbl:lowz_samples}\tablecolumns{6}\tablewidth{0pc}
\tablecaption{Low-redshift volume-limited samples of galaxies selected
  by luminosity  and redshift}  \tablehead{ \colhead{} &  \colhead{} &
  \colhead{} &  \colhead{} & \multicolumn{2}{c}{Sub-shell information}
  \\  \cline{5-6}  \colhead{\sc  Sample} &  \colhead{$M_{^{0.1}r}$}  &
  \colhead{$z$}  &  \colhead{Num.  of  galaxies}  & \colhead{Num.   of
    sub-shells}  &   \colhead{$\Delta{z}_{sub}$}  \\  \colhead{(1)}  &
  \colhead{(2)}  &  \colhead{(3)} &  \colhead{(4)}  & \colhead{(5)}  &
  \colhead{(6)} } \startdata VL1 & (-21.0,-20.2) & (0.03,0.07) & 20553
&  10 &  0.004\\  VL2 &  (-23.0,-21.0) &  (0.03,0.07)  & 5879  & 10  &
0.004\\ VL3 & (-21.0 -20.2) & (0.07,0.11)  & 63694 & 5 & 0.008\\ VL4 &
(-23.0 -21.0) & (0.07,0.11) & 19138  & 5 & 0.008\\ VL5 & (-23.0,-21.0)
& (0.11,0.15) & 40226 & 4 & 0.01\\ VL6 & (-23.0,-21.9) & (0.11,0.15) &
2330 &  4 &  0.01\\ VL7  & (-23.0,-21.9) &  (0.15,0.19) &  3935 &  4 &
0.01\\ VL8 & (-23.0,-21.9) & (0.19,0.23) & 5268 & 4 & 0.01\\ \enddata
\end{deluxetable}

\figurenum{1}

\figurenum{2}

\figurenum{3}

\figurenum{4}

\figurenum{5}

\figurenum{6}

\figurenum{7}

\figurenum{8}

\figurenum{9}

\figurenum{10}

\figurenum{11}
\begin{figure}
\label{fig:profile_lowz}\epsscale{0.9}\plotone{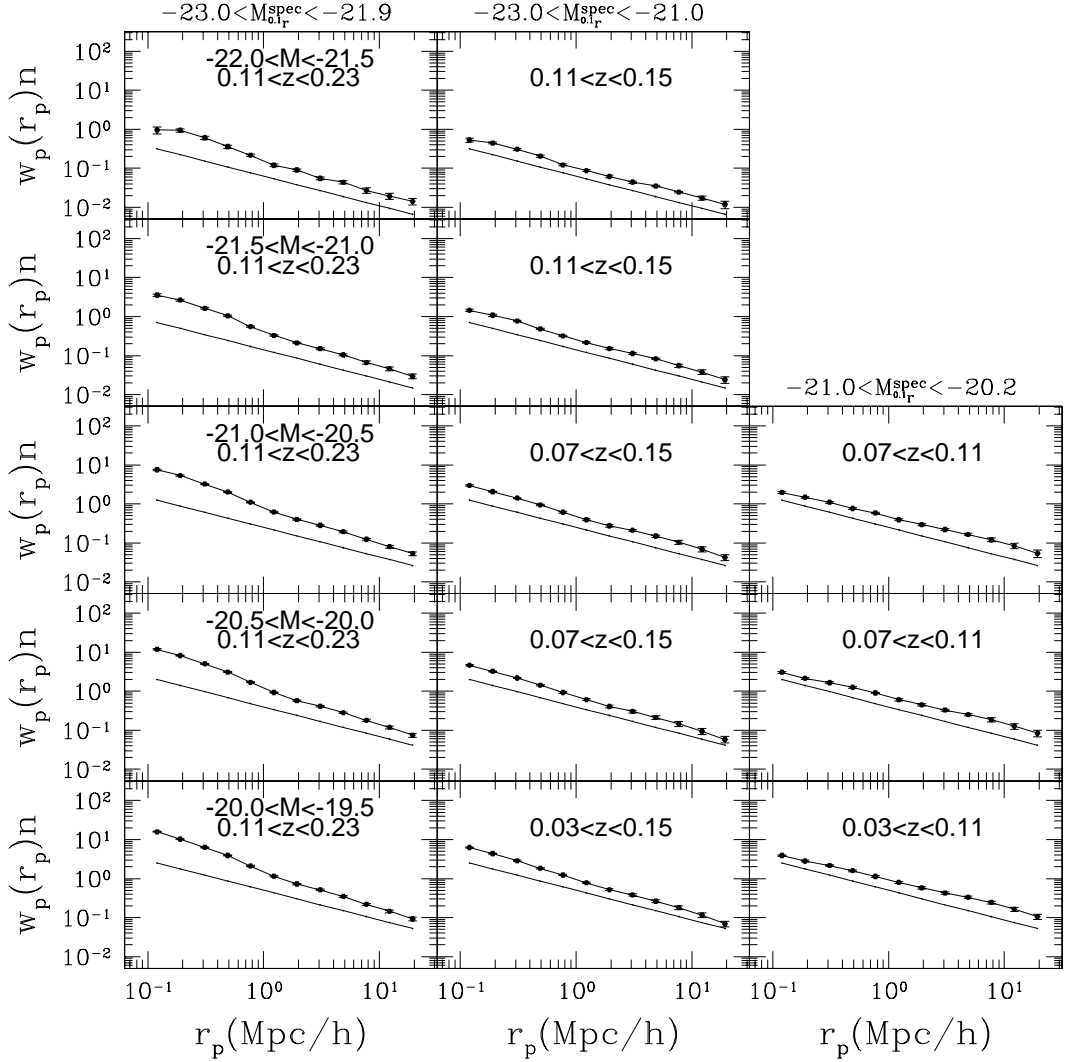}
\caption{Projected density profile $w_p(r_p)n$ between
spectroscopic galaxies of different luminosities (indicated above each
column) and photometric galaxies of different luminosities and redshifts
(both indicated in each panel). A solid black line is repeated on every
panel to guide the eye.}
\end{figure}

\figurenum{11}
\begin{figure}
\epsscale{0.9}\plotone{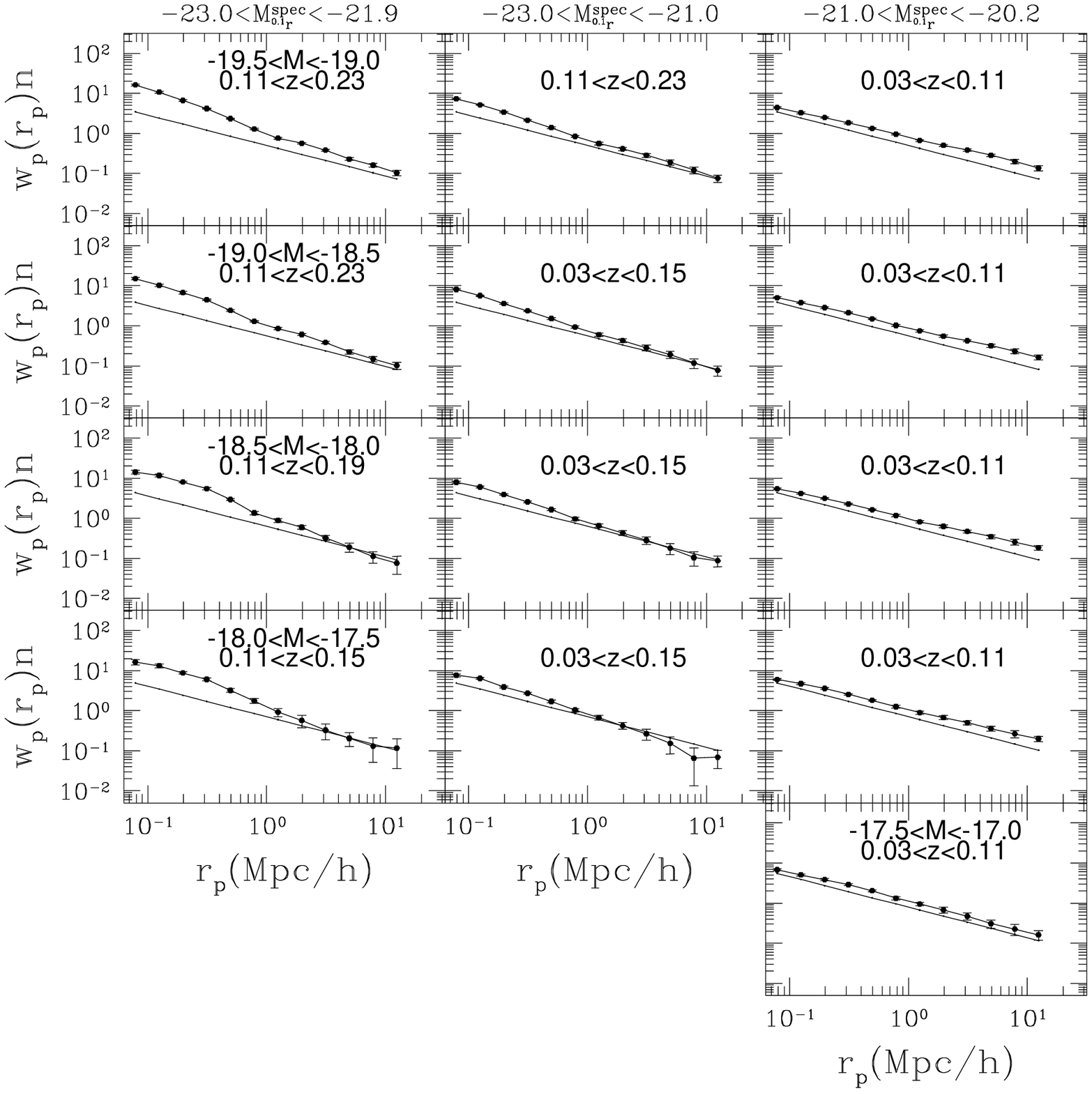}
\caption{Continued...}
\end{figure}

\figurenum{12}
\begin{figure}
\label{fig:bias_lowz}\epsscale{0.9}\plotone{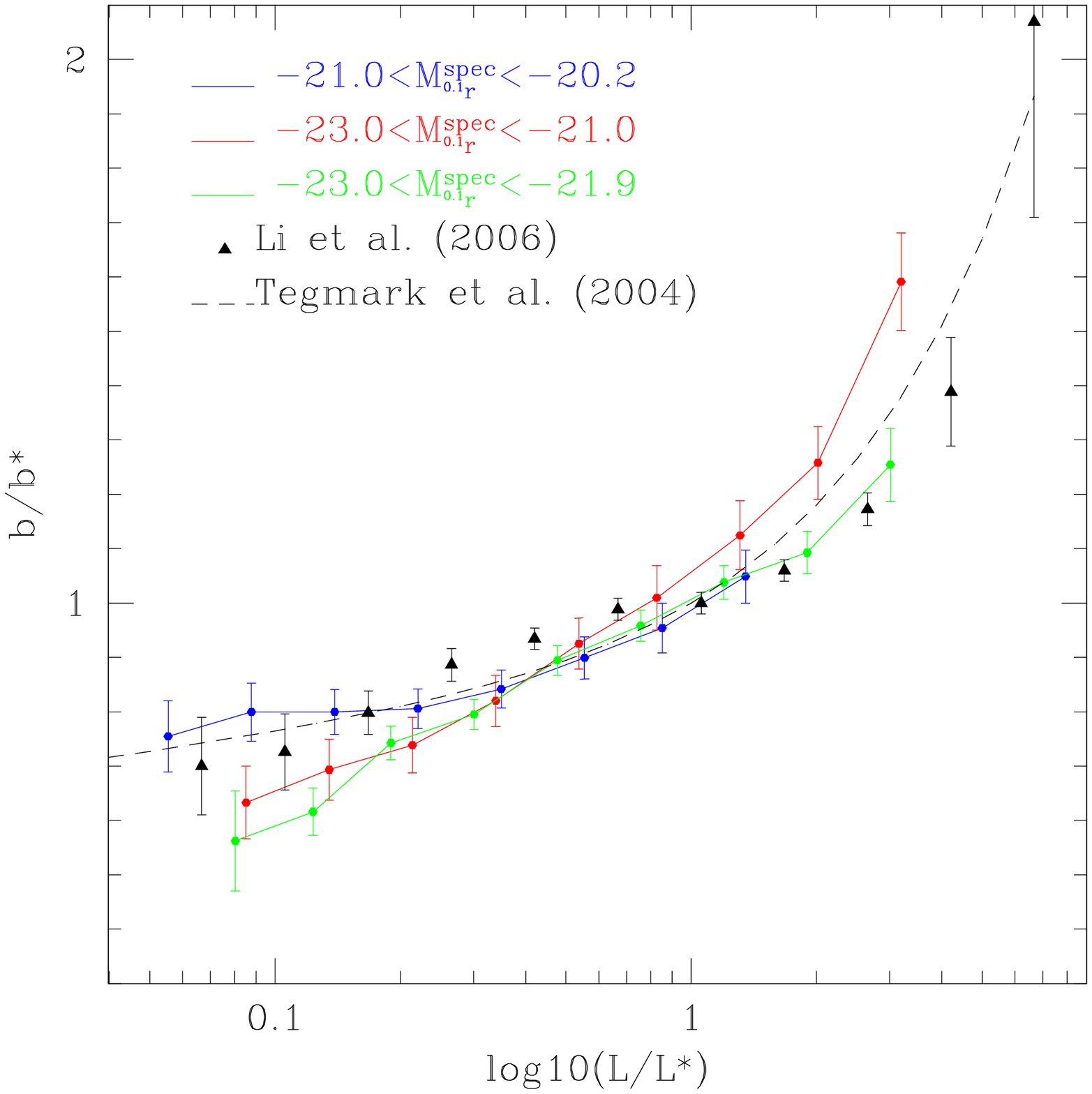}
\caption{Relative bias factor as a function of luminosity, measured for
low-redshift samples. Symbols and lines are similar as in Figure~10.}
\end{figure}

\end{document}